\documentclass[11pt]{article} 

\usepackage{amsmath,amsfonts,bm}

\def\eqref#1{equation~\ref{#1}}

\def\1{\bm{1}}

\DeclareMathAlphabet{\mathsfit}{\encodingdefault}{\sfdefault}{m}{sl}
\SetMathAlphabet{\mathsfit}{bold}{\encodingdefault}{\sfdefault}{bx}{n}

\usepackage{natbib}
\usepackage{hyperref}
\usepackage{url}
\usepackage{subfig}
\usepackage{graphicx}
\usepackage{multirow}
\usepackage{listings}
\usepackage{xcolor}
\usepackage{array}
\usepackage{listings}    
\date{}

\title{\bf Conditioning Trick for Training Stable GANs}
\author{
Mohammad Esmaeilpour, Raymel Alfonso Sallo, Olivier St-Georges\\
Patrick Cardinal, Alessandro Lameiras Koerich\\
\'{E}cole de Technologie Sup\'{e}rieure (\'{E}TS)\\
Universit\'{e} du Qu\'{e}bec\\
Montreal, QC, Canada \\
}

\begin{document}

\maketitle
\begin{abstract}
In this paper we propose a conditioning trick, called difference departure from normality, applied on the generator network in response to instability issues during GAN training. We force the generator to get closer to the departure from normality function of real samples computed in the spectral domain of Schur decomposition. This binding makes the generator amenable to truncation and does not limit exploring all the possible modes. We slightly modify the BigGAN architecture incorporating residual network for synthesizing 2D representations of audio signals which enables reconstructing high quality sounds with some preserved phase information. Additionally, the proposed conditional training scenario makes a trade-off between fidelity and variety for the generated spectrograms. The experimental results on UrbanSound8k and ESC-50 environmental sound datasets and the Mozilla common voice dataset have shown that the proposed GAN configuration with the conditioning trick remarkably outperforms baseline architectures, according to three objective metrics: inception score, Fr\'{e}chet inception distance, and signal-to-noise ratio. 
\end{abstract}

\section{Introduction}
Generative models have been widely used in several audio and speech processing tasks such as speaker verification \citep{reynolds2000speaker}, enhancement \citep{chehrehsa2016speech}, synthesis \citep{raitio2010hmm}, etc. In the last few years, many generative adversarial network (GAN) architectures have been introduced \citep{bollepalli2019generative, sriram2018robust} and they have contributed for tackling such challenging tasks. Furthermore, GANs have been employed in high-level data augmentation for supervised, semi-supervised, and unsupervised audio and speech classification \citep{hu2018generative, donahue2018exploring}. In augmentation with paired transformations, cycle-consistent GANs have been developed for environmental sound classification \citep{esmaeilpour2020unsupervised} as well as for more sophisticated tasks such as voice conversion \citep{fang2018high}. A typical GAN runs a minimax game between generator and discriminator networks, where the latter should distinguish the real sample from the generated example. Following such a baseline GAN, many other variants in terms of similarity metric, loss function, and architectures have been introduced such as the Wasserstein GAN \citep{arjovsky2017wasserstein}, GANs with least squares loss \citep{mao2017least}, attention-based architectures for adversarially training \citep{zhang2019self}, etc. However, training and tuning such advanced GAN models have always been a mounting concern and a difficult challenge due to their instability in training \citep{salimans2016improved, arjovsky2017towards,yang2017statistical}.  

Our main contribution in this paper is in response to this issue. We propose to condition the generator for minimizing dissimilarity among generated and real samples using a normality metric computed in the spectral domain of Schur decomposition \citep{van1983matrix}. This binding improves stability of the generator and does not limitate exploring all the possible modes during training. Without lack of generalizability, our case study in this work is spectrogram synthesis in the domain of audio and speech processing. Spectrograms are 2D representations of 1D signals without phase information \citep{lang1998time}. Spectrogram either from short-time Fourier transformation (STFT) and its variants such as Mel-frequency cepstral coefficients \citep{M5709752} or discrete wavelet transform (DWT) is a standard and common signal representation in signal processing due to its lower dimensionality which densely encodes local frequency information \citep{mallat1999wavelet,rioul1991wavelets}. The highest recognition accuracies have been achieved for classifiers trained on frequency-magnitude representations \citep{hannun2014deep} and this indicates the importance of yielding high quality and informative spectrograms for improving classification performance.

For objectively measuring the impact of this conditioning on the baseline models and our slightly modified BigGAN architecture \citep{brock2018large}, we compute both the inception score (IS) \citep{salimans2016improved} and Fr\'{e}chet inception distance (FID) \citep{heusel2017gans}. Since the quality of the generated spectrograms might not be easily interpretable by human eye, we reconstruct audio signals with some preserved phase information from original samples and measure the signal-to-noise ratio (SNR). Our experimental results on two complete environmental sound and partial Mozilla common voice datasets have shown considerable improvement in generating diverse signals with high fidelity. The rest of the paper is organized as follows. In Section~\ref{sec:rw}, we review some related works and in Section~\ref{sec:conditionGenereator}, we explain theories for constraining the generator. We provide our experimental results and associated discussions in Section~\ref{sec:experiments}. 

\section{Background}
\label{sec:rw}
The generator network $G(\mathbf{z};\theta_{g})$ commonly learns to map from $p_{z}\sim \mathcal{N}(0,I)$ or $\mathcal{U}[-1,1]$ to $p_{g}$ for $\mathbf{z} \in \mathbb{R}^{d_{z}}$ and the discriminator network $D(\mathbf{x}; \theta_{d})$ should maximize $\mathbb{E}_{\mathbf{z}\sim p_{z}(\mathbf{z})}$ $\left[\log\left ( 1-D(G(\mathbf{z}))\right)\right]$ against the generator as defined in (\ref{gan1}) \citep{goodfellow2014generative}.
\begin{equation}
    \min_{G} \max_{D} \mathbb{E}_{\mathbf{x}\sim p_{r}(\mathbf{x})}\left [ \log D(\mathbf{x}) \right ]+\mathbb{E}_{\mathbf{z}\sim p_{z}(\mathbf{z})}\left [ \log \left ( 1-D(G(\mathbf{z})) \right ) \right ]
    \label{gan1}
\end{equation}

The generator $G(\mathbf{z};\theta_{g})$ and discriminator $D(\mathbf{x}; \theta_{d})$ networks are often modeled by convolutional neural networks with different architectures \citep{radford2015unsupervised}. The optimization problem of (\ref{gan1}) not only requires carefully-tuned hyperparameters, but also is very unstable and often collapses during training \citep{salimans2016improved, thanh2019improving}. Generally, there are two approaches to address this issue. First, changing the optimization functions mainly for the generator \citep{fedus2017many,zhang2019self, karras2018progressive, dumoulin2016adversarially, nowozin2016f, salimans2016improved, chen2016infogan, sonderby2016amortised, odena2017conditional}. Likewise, changing the similarity metric from Jensen-Shannon divergence (JSD) to Wasserstein loss \citep{arjovsky2017wasserstein} and Pearson $\chi^{2}$ divergence in least-square GAN \citep{mao2017least} have also been proposed. Second, constraining the discriminator network to provide meaningful gradients everywhere to $G(\mathbf{z};\theta_{g})$ \citep{mescheder2018training, miyato2018spectral, gulrajani2017improved, kodali2017convergence}.

There are two approaches that are relevant to our work for improving stability in GAN training. The first one is focused on bijective mapping between $p_{g}$ and $p_{r}$ using another network such as an autoencoder \citep{donahue2016adversarial}. This ensures a correlation between generated and real samples through a regularization function similar to $ \min \left \| \mathbf{z}-Rec(G(\mathbf{z})) \right \|_{2}+ H(\mathbf{z},Rec(\mathbf{x}))$ where $Rec$ denotes the reconstructor network (autoencoder) and $H$ is the entropy loss \citep{srivastava2017veegan}. Besides, some variational autoencoder schemes have been also proposed for avoiding instability \citep{kingma2014stochastic, rezende2014stochastic}. The effectiveness of explicit regularization of loss functions with autoencoder loss has been studied by \citet{che2016mode}. They have introduced several costly metrics for estimating modes and enhancing quality of the generated samples. A similar metric for encoding random samples from $p_{r}$ to $p_{g}$ has been developed by \citet{larsen2015autoencoding}. Both of these approaches incorporate $\mathcal{L}\left [ \mathbf{x}, G(Rec(\mathbf{x})) \right ]$ which is a pixel-wise loss followed by another regularization term. The second approach is spectral normalization \citep{miyato2018spectral} which conditions the discriminator to support Lipschitz continuity using singular value decomposition. This approach regularizes $\theta_{d}$ (the weight matrix of the discriminator) towards the direction of the top rank (first) singular value. Inspired by \citet{zhang2019self}, it has been shown that this regularization can be more effectively implemented for $G(\mathbf{z};\theta_{g})$ as \citep{brock2018large}:
\begin{equation}
    \theta_{g} = \theta_{g}-\max(0,\sigma_{0}-\sigma_{clamp})v_{0}u_{0}^{\top}
\end{equation}
\noindent where $\theta_{g}$ denotes the weight matrix of the generator and $v_{0}\sigma_{0}u_{0}^{\top}$ forms the first basis matrix of $\theta_{g}$ after decomposition. The threshold $\sigma_{clamp}$ can be set to a predefined value (which implies an additional hyperparameter) or $\sigma_{clamp} = \mathrm{sg}(\sigma_{1})$; where $\mathrm{sg}$ stands for the stop-gradient operation. This normalization can be generalized to other subsequent singular values, efficiently computed by Alrnoldi method \citep{golub2000eigenvalue}.

While instability is a common problem when training GANs with any dataset, it is critical for 2D representations of audio and speech, mainly due to the properties of Fourier or wavelet transforms \citep{esmaeilpour2020unsupervised}. In the next section, we explain how to control the generator for producing high fidelity spectrograms through correctly conditioning it.

\section{Conditioning the Generator}
\label{sec:conditionGenereator}
In light of correlating $p_{g}$ to $p_{r}$ and spectral normalization in response to the instability issue in adversarial training, we condition $G(\mathbf{z};\theta_{g})$ in another spectral domain to enhance the stability and also in order to provide better gradients everywhere to the discriminator. Our conditioning is fundamentally different from employing an autoencoder or regularizing either $\theta_{g}$ or $\theta_{d}$ through normalizing their singular value(s). Instead of normalizing the top rank eigenvalues of $\theta_{g}$ in spectral normalization \citep{zhang2019self}, we asymptotically correlate $p_{g}$ to $p_{r}$ in such a way that it forces the generator to follow Schur spectral distribution $p_{r}$. In fact, we force the generator to get closer to the departure from normality function computed for the original spectrograms in $p_{r}$ defined in the spectral domain of the generalized Schur decomposition \citep{van1983matrix}. This binding makes the generator amenable to the truncation trick \citep{brock2018large} and the discriminator might converge in fewer iterations. Additionally, it neither limits spanning the entire possible modes nor loses sample variety. 

\textbf{Lemma.} For an input sample (spectrogram) $\mathbf{x}_{i}$ from a given distribution, there exists a unitary representation $Q \in \mathbb{C}^{n \times n}$ in such a way that \citep{golub2012matrix}:
\begin{equation}
    Q^{H}\mathbf{x}_{i}Q = V+S
    \label{eq:schurGener}
\end{equation}
\noindent where $Q^H$ denotes the conjugate transpose of $Q$ in vector space of Schur decomposition, $S=\left \{ s_{i}\mid i=0:n-1 \right \} \in \mathbb{C}^{n \times n}$ is an upper triangular matrix, and $V=\mathrm{diag}(\lambda_{0},\lambda_{1},\cdots$ $,\lambda_{n-1})$ contains eigenvalues of $\mathbf{x}_{i}$ ($\lambda_{i}$ denotes an independent eigenvalue for $\mathbf{x}_{i}$). In this unitary vector space, which might also yield a quasi-upper triangular representation for $S$, $Q=\left [ q_{0}\mid  q_{1} \mid  q_{2} \mid  \cdots \mid  q_{n-1} \right ]$ provides the pencil of $\overrightarrow{q_{i}}-\lambda_{i} \overrightarrow{q_{i+1}}$ for $i\leq n-2$ which is also known as basis vector for $\mathbf{x}_{i}$. According to the perpendicularity of the achieved pencils and the support matrix $S$, we write:
\begin{equation}
    \mathbf{x}_{i}q_{k}\approx \lambda_{k}q_{k}+\sum_{i=0}^{n-1}s_{ik}q_{i}, \quad k=0:n-1.
    \label{basisfunc}
\end{equation}
\noindent Therefore with the general assumption of quasi-upper triangular subspaces with the normal span of $\left \{ q_{0},q_{1}, \cdots, q_{k} \right \}$ for $k=0:n-1$, we can conclude that the choice of $S$ should be independent of $Q$ \citep{golub2012matrix}. Accordingly, we can compute its Frobenius norm using $\lambda_{i}$ as:
\begin{equation}
    \left \| S \right \|_{F}^{2} = \left \| \mathbf{x}_{i} \right \|_{F}^{2}-\sum_{i=0}^{n-1}\left | \lambda_{i} \right |^{2}\equiv \Delta^{2} (\mathbf{x}_{i})
    \label{f_gan2}
\end{equation}
\noindent where $\Delta^{2}$ is known as departure from normality (DFN) \citep{golub2012matrix}. 
For ensuring the correlation between $\mathbf{x}_{r}$ and $\mathbf{x}_{g}$ randomly drawn from $p_{r}$ and $p_{g}$ in their designated vector spaces (span of $q_{i}$s), the DFN metric should support $\left | \Delta^{2}(\mathbf{x}_{j}) - \Delta^{2}(\mathbf{x}_{i}) \right |<\epsilon$ for a small enough $\epsilon$. Such a DFN condition also ensures the consistency of corresponding pencils and contributes to the generalized form of Schur decomposition with $Q_{k}^{H}\mathbf{x}_{r}Z_{k}=R_{k}(V_{k}+S_{k})$ and $Q_{k}^{H}\mathbf{x}_{g}Z_{k}=R_{k}(V_{k}+S_{k})$, where $R_{k}=Q_{k}^{H}(\mathbf{x}_{r}\mathbf{x}_{g_{k}}^{-1}Q_{k})$ and $Z_{k}$ is also unitary and supports for $\lim_{i\rightarrow \infty}(Q_{k_{i}}, Z_{k_{i}})=(Q,Z)$ \citep{golub2012matrix}. 
The intuition behind exploiting these basis vectors is providing pencils of  $\overrightarrow{\mathbf{x}_{r_{i}}}-\lambda_{i}\overrightarrow{\mathbf{x}_{g_{i}}}$ for learning the original distribution $p_{r}$ by the generator. The derivable pencils are not necessarily normal in the span of their associated subspaces, however, their linear combination imparts $p_{r}$ to $p_{g}$ (in the closed form). Furthermore, diagonal values in $V$ constitutes coefficient of basis vectors (sub-pencils in their manifolds) and represent local properties of the given input sample.

\textbf{Proposition.} The DFN metric in the form $\left | \Delta^{2}(\mathbf{x}_{g}) - \Delta^{2}(\mathbf{x}_{r}) \right |<\epsilon$ ensures the correlation of $\mathbf{x}_{g} \sim p_{g}$ and $\mathbf{x}_{r} \sim p_{r}$ in the spectral domain with a measurable error term $\epsilon$.

\textbf{Proof.} Since $\Delta^{2}(.)$ is differentiable in its designated subspaces, we can find an upper bound for $\epsilon$. For all $x_{i}:=\max(\mathrm{diag}(S_{i}))$ we assume $g_{r}(x_{i})=\Delta^{2}(\mathbf{x}_{i}) \in \mathbb{C}^{n+1}$ with degree $n+1$ and $g_{g}(x_{i})=\Delta^{2}(G(\mathbf{z}_{i})) \in \mathbb{C}^{n}$ with degree $n$ are differentiable over the interval $[\varpi_{inf},\varpi_{sup}]$, therefore we can approximate the error function as \citep{phillips2003interpolation}:
\begin{equation}
   e(x)=g_{r}(x)-g_{g, n}(x)=\frac{g^{(n+1)}_{r}(\xi)}{(n+1)!}\prod_{i=0}^{n}(x-x_{i})
    \label{errorfunc}
\end{equation}
\noindent where $\xi \in (\varpi_{inf},\varpi_{sup})$ with the marginal condition using the second derivative $\left | \ddot{g}_{r}(x) \right |<\varrho$ for $0 \leq \varrho <<1$ and $n=1$ we write: 

\begin{multline}
    g_{r}(x)-g_{g,1}(x)=\underbrace{\left ( x-x_{i} \right )\left ( x-x_{i+1} \right )}_{g_{f}(x)}\frac{\ddot{g}_{r}(\xi)}{2!} \Rightarrow \dot{g}_{f}(x)=2x-(x_{i}+x_{i+1})=0\Rightarrow  x=\frac{x_{i}+x_{i+1}}{2} \\ g_{f}\left ( \frac{x_{i}+x_{i+1}}{2} \right )=-\left ( \frac{x_{i+1}-x_{i}}{2} \right )^{2} \Rightarrow
    \left | e(x) \right |\leq \frac{\ddot{g}_{r}(\xi)}{8}\left ( x_{i+1}-x_{i} \right )^{2} \quad \square
\end{multline}

Herein, $\left | e(x) \right |$ measures how the dynamic $\Delta_{g}^{2}(\cdot)$ is close to the static $\Delta_{r}^{2}(\cdot)$ and equivalently measures the correlation between $p_{g}$ and $p_{r}$. We experimentally demonstrate that minimizing the generator $G(\mathbf{z})$, subject to:
\begin{equation}
    \left | \mathbb{E}_{\mathbf{z}\sim p_{z}(\mathbf{z})} \Delta^{2}(G(\mathbf{z})) - \mathbb{E}_{\mathbf{x}\sim p_{r}(\mathbf{x})} \Delta^{2}(\mathbf{x}) \right |<\epsilon
    \label{eq:conditionGenet}
\end{equation}

\noindent yields high quality spectrograms and improves stability. 

\subsection{Fast DFN Approximation}
Computing $\sum_{i=0}^{n-1}\left | \lambda_{i} \right |^{2}$ as defined in (\ref{f_gan2}) can be computationally prohibitive since it involves recursive multiplication of pencils. In response to this issue, $f_{V}(\lambda_{i})$ can be defined as a polynomial function originating from the absolute product of eigenvalues \citep{edelman1993circular,van1983matrix}:
\begin{equation}
   \sum_{i=0}^{n-1}\left | \lambda_{i} \right |^{2}\cong  \sum_{i=0}^{n-1} \left ( \prod_{i=0}^{n-1} f_{V}(\lambda_{i})\prod_{i\neq j}^{n-1} \left | \lambda_{i} - \lambda_{j} \right |\underbrace{e^{-\sum \lambda_{i}^{2}/2}}_{\text{Normal distribution}} \right )
   \label{eq:approxEign}
\end{equation}
\noindent where it imparts a normal distribution and closely relates to the joint density function of the eigenvalues for a random symmetric matrix $\mathbf{\mho}_{k\times k} :=(\varrho +\varrho^{\top})/2$. We derive $\varrho$ by downsampling a given input spectrogram (i.e., $\mathbf{x}_{n\times n}$) by a factor of two. Every single element of the derived Toeplitz matrix ($\mho$) should be also distributed according to a normal distribution \citep{van1983matrix}. The probability distribution of $\mho$ which is also known as Gaussian orthogonal ensemble \citep{alt1995gaussian,wu1990gaussian} for the eigenvalues of $\lambda_{i}$ has the form of \citep{edelman1993circular}:
\begin{equation}
    \frac{2^{-k/2}}{\prod_{i=1}^{k}\nu (i/2)}\prod_{i\neq j}^{k-1} \left | \lambda_{i} - \lambda_{j} \right |e^{-0.5\sum \lambda_{i}^{2}}
\end{equation}
\noindent where $\nu(\cdot)$ denotes Minkowski function \citep{panti2008multidimensional,thompson1996minkowski}. Accordingly, the summation of (\ref{eq:approxEign}) for fast approximation (\ref{f_gan2}) is the expectation over the determinants of $\mathbf{\mho}$ as \citep{edelman1993circular}:
\begin{equation}
    \frac{1}{k!}\sum_{i=0}^{n-1} \left ( \prod_{i=0}^{n-1} f_{V}(\lambda_{i})\prod_{i\neq j}^{n-1} \left | \lambda_{i} - \lambda_{j} \right |e^{-\sum \lambda_{i}^{2}/2} \right ) \approx   \left \{ 2^{k/2}\prod_{i=0}^{k-1}\nu(i/2)  \right \}\mathbb{E}_{\mho}\det\left ( \underbrace{\mho}_{f_{V}(\lambda_{i})} \right )
\end{equation}

\noindent Although there are some analytical approaches for closely approximating $f_{V}(\lambda_{i})$, we simply use $\mho$ since it is normally distributed over $\mathbf{x}$ and $\lambda_{i}\rightarrow 0$ for $i>0.5 \cdot k$. Therefore $\det(\mho)$ gives an acceptable approximation.

\section{Experimental Results}
\label{sec:experiments}
In this section, we provide details of our experiments on two benchmarking environmental sound datasets: UrbanSound8K \citep{Salamon:UrbanSound:ACMMM:14} and ESC-50 \citep{piczak2015esc}. We have also conducted a brief ablation study for Mozilla common voice (MCV)\footnote{\url{https://voice.mozilla.org/en/datasets}}. The first two datasets include 8,732 and 2,000 short environmental audio signals ($\leq$5 sec) organized in 10 and 50 different classes, respectively. MCV consists of 4,257 recorded hours of multi-language speeches ($\leq$7 sec) and the corresponding text transcriptions. For generating DWT spectrograms, we use complex Morlet mother function with static sampling frequency of 16 kHz and frame length of 50 ms with 50\% overlapping. We represented each spectrogram using three different visualizations of linear, logarithmic, and logarithmic-real magnitude scales (see Appendix~\ref{appendix:GSpecs})\citep{esmaeilpour2020sound}. For STFT spectrograms, we set the number of filters to 2,048 with the hop length of 1,024 and reflect padding in overlapping audio chunks of 50 ms (with ratio 0.5) \citep{esmaeilpour2020sound}. Meanwhile, we apply pitch shifting \citep{salamon2017deep} with scales $0.75, 0.9, 1.15,$ and $1.5$ for audio signals with length $\leq 2$ seconds for data augmentation \citep{esmaeilpour2019robust}.

Our baseline model is the SA-GAN architecture \citep{zhang2019self} using the hinge loss objective for smooth training \citep{lim2017geometric}. Since we employ class-conditional (CC) learning similar to \citet{mirza2014conditional}, the generator receives CC batch-norm as suggested by \citet{de2017modulating}. For the discriminator, we run the identical procedure with projection \citep{miyato2018cgans} following the same optimization procedure in the baseline SA-GAN, however with some modifications in the number of channels and applying spectral regularization for the generator network. We have evaluated different options for initializing both networks, such as Glorot \citep{glorot2010understanding} and variants of $\mathcal{N}(0,I)$ (e.g., $\mathcal{N}(0,0.02I)$) \citep{RadfordMC15}. Since this choice noticeably affects the training performance \citep{brock2018large} and requires to be evaluated individually according to properties of the dataset, we eventually used orthogonal regularization \citep{SaxeMG13}.

\begin{figure}[htpb!]
  \centering
  \includegraphics[width=\textwidth]{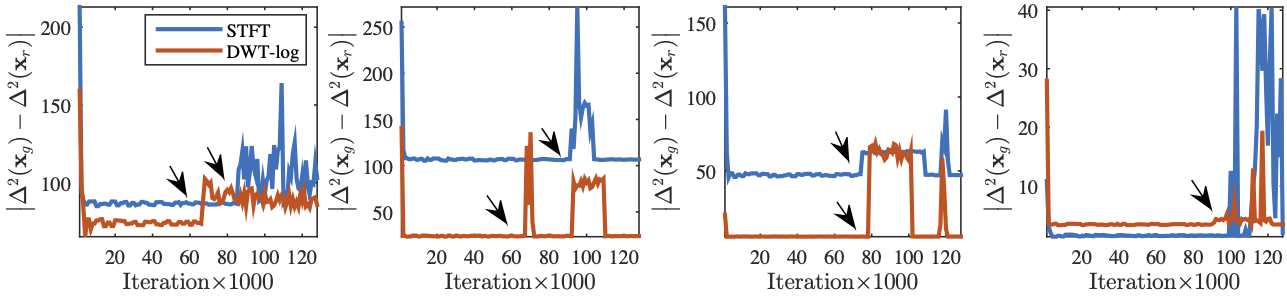}
\begin{tabular}{p{0.1\textwidth}p{0.22\textwidth}p{0.22\textwidth}p{0.22\textwidth}p{0.22\textwidth}}
  & \scriptsize (a) & \scriptsize (b) & \scriptsize (c) & \scriptsize (d)
  \end{tabular}
  \vspace{-15pt}
  \caption{The typical plot of the difference DFN measures for a random non-convolution layer in $G(\mathbf{z};\theta_{g})$ trained on STFT and logarithmic DWT representations of UrbanSound8K dataset. (a) The baseline SA-GAN, (b) the BigGAN architecture with spectral normalization in both $G(\mathbf{z};\theta_{g})$ and $D(\mathbf{x}; \theta_{d})$, (c) the BigGAN configuration with spectral normalization only in $D(\mathbf{x}; \theta_{d})$, and (d) the scaled up (by a factor of 2) configuration for (c). Arrows in these sub-figures refer to collapse onsets.} 
  \label{gplots}
\end{figure}

Figure~\ref{gplots} depicts the performance of four GAN architectures trained on logarithmic DWT spectrograms. All these models undergo collapse at different iterations which forces to early stop the training. All the subsequent quality measurements on the performance of these models will be conducted on the checkpoints prior to collapse. We compute the $\left | \Delta^{2}(\mathbf{x}_{g}) - \Delta^{2}(\mathbf{x}_{r}) \right |$ for each model to demonstrate collapse without imposing the difference DFN condition on their generators (see inequality~(\ref{eq:conditionGenet})). These plots corroborate the positive impact of spectral normalization on delaying collapse at further iterations. The best performance of this normalization is when it is applied only on $D(\mathbf{x}; \theta_{d})$ (see Figures~\ref{gplots}(b) and~\ref{gplots}(c)), though incorporating it into both $G(\mathbf{z};\theta_{g})$ and $D(\mathbf{x}; \theta_{d})$ still outperforms the baseline SA-GAN.

Furthermore, we have evaluated the impact of increasing the batch size on the performance of GANs during training while monitoring the difference DFN measure along iterations. We scaled up the batch size to 256 and 512 aiming at covering more modes and providing better gradients to both $G(\mathbf{z};\theta_{g})$ and $D(\mathbf{x}; \theta_{d})$ \citep{brock2018large}. Figure~\ref{gplots}(d) depicts this effect for the BigGAN with batch size of 512 and regularized $\theta_{d}$ using spectral normalization. According to this graph although this GAN keeps its stability until about 100k iterations, it undergoes complete collapse afterward. Moreover, the difference DFN measure is considerably improved compared to other three GANs which indicates convergence in fewer iterations. We additionally scaled up the number of channels (width) in each layer of the networks by a factor of 2 at the cost of doubling the number of required parameters. However this did not rectify nor delay the collapse at higher iterations.  

We slightly modify the BigGAN architecture for class conditional (CC) training (see Appendix~\ref{appendix:A}). Three major settings in CC training are required to adapt our benchmarking datasets. Firstly setting the conditional vector ($c$-embedding) and secondly, the skip connection (direct skip-$z$) from the given noise vector according to the designated probability distribution. While using separate layers for $c$-embeddings has been proposed by \citet{miyato2018spectral}, we also found shared embeddings \citep{PerezSVDC18, brock2018large} outperforms the latter. For the skip-$z$ connection probe, we firstly evaluated splitting $z \sim \mathcal{N}(0,I)$ and $z \sim \mathcal{N}(0,0.02I)$  vectors into smaller chunks (we set to 20) and concatenating them into $c$-embeddings in each level of resolution as suggested by \citet{brock2018large}, then explored a couple of other variants \citep{denton2015deep, goodfellow2014generative}. We ended up to concatenate the vector $z$ in its entire dimension to $c$-embeddings \citep{brock2018large}. Finally, the third setting is the number of units in each hidden layer of the networks (the channel multiplier). Increasing this hyperparameter relatively affects the IS and FID scores for the model.

Our objective evaluations are summarized in Table~\ref{table:Comp1} and the number of iterations without collapse indicate considerable stability improvement compared to baseline GANs shown in Figure~\ref{gplots}. This table also compares the effect of batch size and number of channels on the performance of our slightly modified BigGAN using spectral normalization and difference DFN measure. According to this table, doubling the batch size considerably improves both the IS and FID. Among different configurations for the GAN, models trained on three visualizations of DWT spectrograms dominantly outperform the STFT. We conjecture that this is due to the complexity of Morlet mother function \citep{young2012wavelet} compared to sinusoidal transform in STFT.

\begin{table}[t]
\centering
\footnotesize
\begin{tabular}{c|c c c c}
\hline
Dataset               & \begin{tabular}[c]{@{}c@{}}Orthogonal\\ Regularization\end{tabular} & \begin{tabular}[c]{@{}c@{}}Iteration\\ $\times 1000$\end{tabular}         & IS                                                                                                                            & FID                                                                                                                                  \\ \hline \hline
\multirow{4}{*}{US8K} & \begin{tabular}[c]{@{}c@{}}False\\ ($batch=256$)\end{tabular}       & \begin{tabular}[c]{@{}c@{}}$132$\\ $\left ( ch.=64 \right )$\end{tabular} & \begin{tabular}[c]{@{}c@{}}$43.12\pm 1.46$\\ $ \left (\downarrow 3.50, \uparrow 1.148, - \right )$\end{tabular} & \begin{tabular}[c]{@{}c@{}}$44.57\pm 2.93$\\ $\left (\uparrow 1.67, \downarrow 2.95, \uparrow 22.23 \right )$\end{tabular}          \\ \cline{2-5} 
                      & \begin{tabular}[c]{@{}c@{}}True\\ ($batch=256$)\end{tabular}        & \begin{tabular}[c]{@{}c@{}}$138$\\ $\left ( ch.=96 \right )$\end{tabular} & \begin{tabular}[c]{@{}c@{}}$46.89 \pm 0.12$\\ $\left (\downarrow 2.64, \uparrow 1.54, - \right )$\end{tabular}  & \textbf{\begin{tabular}[c]{@{}c@{}}$29.06\pm 1.18$\\ $\left (\uparrow 5.28, \downarrow 6.71, - \right )$\end{tabular}} \\ \cline{2-5} 
                      & \begin{tabular}[c]{@{}c@{}}False\\ ($batch=512$)\end{tabular}       & \begin{tabular}[c]{@{}c@{}}$134$\\ $\left ( ch.=64 \right )$\end{tabular}  & \begin{tabular}[c]{@{}c@{}}$49.61 \pm 0.12$\\ $\left (\downarrow 2.41, \uparrow 1.78, - \right )$\end{tabular}  & \begin{tabular}[c]{@{}c@{}}$\mathbf{21.75\pm 0.31}$\\ $\left (\downarrow 2.09, \downarrow 3.14, - \right )$\end{tabular}        \\ \cline{2-5} 
                      & \begin{tabular}[c]{@{}c@{}}True\\ ($batch=512$)\end{tabular}        & \begin{tabular}[c]{@{}c@{}}$133$\\ $\left ( ch.=96 \right )$\end{tabular}  & \begin{tabular}[c]{@{}c@{}}$\mathbf{52.36 \pm 1.19}$\\ $\left (\uparrow 1.94, \uparrow 2.72, \downarrow 27.19 \right )$\end{tabular}    & \begin{tabular}[c]{@{}c@{}}$21.93\pm 1.11$\\ $\left (\downarrow 1.40, -, \uparrow 14.18 \right )$\end{tabular}        \\ \hline \hline
\multirow{4}{*}{ESC-50}  & \begin{tabular}[c]{@{}c@{}}False\\ ($batch=256$)\end{tabular}       & \begin{tabular}[c]{@{}c@{}}$148$\\ $\left ( ch.=64 \right )$\end{tabular}  & \begin{tabular}[c]{@{}c@{}}$68.76\pm 3.14$\\ $\left (\uparrow 0.64, \uparrow 5.63, - \right )$\end{tabular}    & \begin{tabular}[c]{@{}c@{}}$34.61\pm 2.23$\\ $\left (\downarrow 2.16, \uparrow 1.77, - \right )$\end{tabular}            \\ \cline{2-5} 
                      & \begin{tabular}[c]{@{}c@{}}True\\ ($batch=256$)\end{tabular}        & \begin{tabular}[c]{@{}c@{}}$143$\\ $\left ( ch.=96 \right )$\end{tabular}  & \begin{tabular}[c]{@{}c@{}}$71.94\pm 1.42$\\ $\left (-, \uparrow 7.91, \downarrow 19.68 \right )$\end{tabular}     & \begin{tabular}[c]{@{}c@{}}$29.81\pm 3.15$\\ $\left (\uparrow 3.62, \downarrow 2.13, - \right )$\end{tabular}          \\ \cline{2-5} 
                      & \begin{tabular}[c]{@{}c@{}}False\\ ($batch=512$)\end{tabular}       & \begin{tabular}[c]{@{}c@{}}$139$\\ $\left ( ch.=64 \right )$\end{tabular}  & \begin{tabular}[c]{@{}c@{}}$71.46\pm 0.26$\\ $\left (\downarrow 1.47, \downarrow 2.38, - \right )$\end{tabular}   & \begin{tabular}[c]{@{}c@{}}$29.03\pm 1.05$\\ $\left (\downarrow 0.89, \downarrow 4.63, -\right )$\end{tabular}        \\ \cline{2-5} 
                      & \begin{tabular}[c]{@{}c@{}}True\\ ($batch=512$)\end{tabular}        & \begin{tabular}[c]{@{}c@{}}$145$\\ $\left ( ch.=96 \right )$\end{tabular}  & \begin{tabular}[c]{@{}c@{}}$\mathbf{73.69\pm 1.47}$\\ $\left (\uparrow 3.55, \uparrow 1.95, - \right )$\end{tabular}     & \begin{tabular}[c]{@{}c@{}}$\mathbf{23.03\pm 2.47}$\\ $\left (\uparrow 2.79, \uparrow 1.31, \uparrow 21.84 \right )$\end{tabular}            \\ \hline
\end{tabular}
\caption{Comparison of the inception score (higher is better) and Fr\'{e}chet inception distance (lower is better) for our proposed modification to BigGAN trained on logarithmic DWT spectrograms (resolution $128 \times 128$) by running spectral normalization and difference DFN minimization on $D(\mathbf{x}; \theta_{d})$ and $G(\mathbf{z};\theta_{g})$, respectively. Scores are averaged over 5k generated spectrograms over 10 different runs for each dataset. Values inside parenthesis in the last two columns correspond to linear-DWT, logarithmic-real-DWT, and STFT representations. Accordingly, up-/down-ward arrows denote increase/decrease in scores relative to the logarithmic DWT. The best results are shown in boldface.}
\label{table:Comp1}
\end{table}

\subsection{Orthogonal Regularization vs. Difference DFN}
The orthogonal regularization \citep{brock2017neural} has been proposed for smooth training, which to some extent, makes a better trade-off between sample variety and fidelity by constraining over ${\theta}_{d}$ as:
\begin{equation}
    R_{\beta}=\beta\left \| \theta_{d}^{\top} \theta_{d}-I \right \|^{2}_{F}
    \label{eq:ortho}
\end{equation}
\noindent where $\beta$ is a small hyperparameter. This regularization prohibitively affects the generator and results to early collapse at about 25k iterations while trained on logarithmic DWT spectrograms. Recently an improved version of this regularization has been introduced \citep{brock2018large}:
\begin{equation}
    R_{\beta}=\beta\left \| \theta_{d}^{\top} \theta_{d}\odot (\mathbf{1}-I)\right \|^{2}_{F}
    \label{eq:uportho}
\end{equation}
\noindent where $\mathbf{1}$ refers to a matrix with constant values of 1. This regularization term binds to minimize local similarity among filters. In our experiment we found out $\beta \in (10^{-4},10^{-5}]$ is the best range for STFT and DWT spectrograms except for the logarithmic-real. Upon running several exploratory experiments, we set $\beta=10^{-4}$ for it. Unlike orthogonal regularization, the difference DFN does not involve weight matrix manipulations at each layer. This considerably reduces the computational cost for training and makes the discriminator converge in fewer iterations and remarkably delays the collapse at higher iterations. Table \ref{table:Comp1} shows the positive effect of orthogonal regularization while used with minimized DFN measure.

\subsection{Spectrogram Fidelity and variety}
\begin{figure*}[htpb!]
  \centering
  \includegraphics[width=\textwidth]{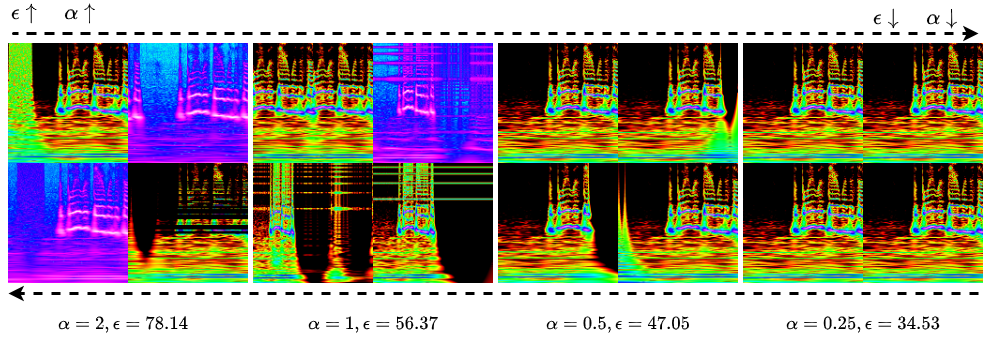}
  \caption{Generated logarithmic DWT spectrograms using our modified BigGAN with spectral normalization and minimized DFN measure.} 
  \label{specplots}
\end{figure*}

Similar to \citet{brock2018large}, our modification to the BigGAN makes the model amenable to truncation for $p_{r} \sim \mathcal{N}(0,\alpha I)$ where the threshold $\alpha \in [0, 1)$. Smaller values for $\alpha$ negatively affects the total number of spectrogram variety however, considerably boosts the quality of the generated samples. This trade-off (see a relevant study in \citet{marchesi2017megapixel}) also affects the actual value of $\epsilon$ in computing difference DFN measure (inequality~(\ref{eq:conditionGenet})). Averaged over different experiments on logarithmic DWT representations, for $\alpha \leq 0.5$ we achieve $\epsilon \leq 37.16$. This indicates higher quality of the generated spectrogram at the cost of losing sample variety. Likewise, increasing $\epsilon$ expands the sample variety but with degraded sample quality (see Figure~\ref{specplots}). Our GAN architecture for $\alpha \leq 0.2$ generates oversmoothed spectrograms as a result of a poorly conditioned model and increases (explodes) $\epsilon$ toward higher values. 

\begin{figure}[htpb!]%
    \centering
    \includegraphics[width=1\textwidth]{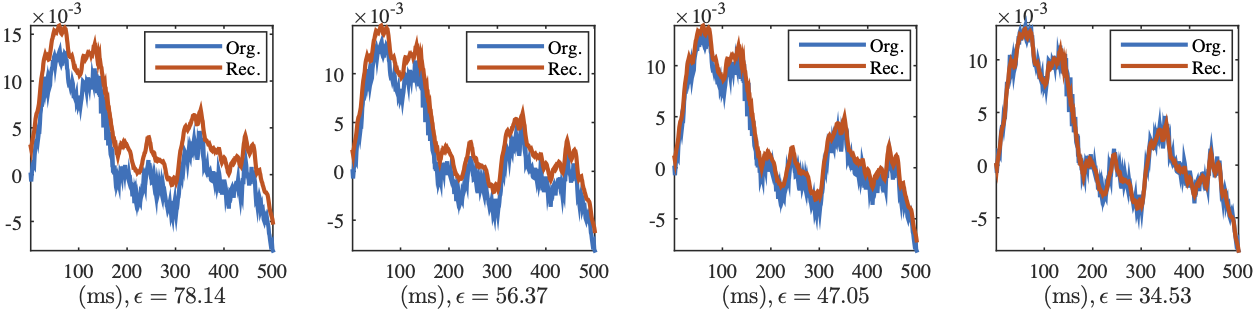}
\caption{Comparison of original and reconstructed sounds from the generated logarithmic DWT spectrograms with identical phase matrix randomly drawn from an original speech signal from MCV. For better visualization we have shown the first 500 ms from the entire three second-length signal.}
\label{graph:rec}
\end{figure}

The truncation trick directly increases the IS and specifically for $\alpha=0.25$ we noticed 12.3\% improvement in generating high quality logarithmic DWT spectrograms compared to $\alpha=2$ or $\mathcal{U}[-1,1]$. Moreover, a slight reduction in $\alpha$ positively impacts the FID by reducing the score 10\% on average, however, when $\alpha\rightarrow 0$ the FID sharply increases. Whereas FID, the IS does not penalize sample variety, however, it rewards precision and this turns out to be a biased objective evaluation on the quality of generated spectrograms \citep{brock2018large}. Moreover, interpretation and quality analysis of the generated spectrograms could be very difficult to human eyes. To address this issue, we reconstruct the audio signals from them and measure the SNR. This requires access to the phase information for each spectrogram \citep{karl2020} and unfortunately we could not successfully train our model with phase vectors. Although, there are some approaches for phase approximation \citep{leeb1989simultaneous,mulgrew2013stationary} or reconstructing signal almost without phase information \citep{kumar2019melgan} using GAN, we noticed they often introduce noise (ambient, background, hissing, etc.) to the reconstructed signal. We eventually opted to utilize subjective phase vectors from original known samples. Figure~\ref{graph:rec} shows that spectrograms with lower values of $\alpha$ and $\epsilon$ better reconstruct audio signals in terms of quality (see Appendix~\ref{appendix:GSpecs} for additional information). 

\subsection{Summary and Ablation Study}
In this section we compare the best performance of our proposed model with other GANs considering $256\times256$ and $128\times 128$ input resolutions, as shown in Table~\ref{table:Comp2}. We use the ablated model for MCV dataset on 100k randomly selected recordings shorter than five seconds. Moreover, we objectively compare the quality of the reconstructed signals using the SNR\footnote{$\mathrm{SNR}_{db}(\mathbf{x}_{r},\mathbf{x}_{g})=20\log_{10}Pw(\mathbf{x})/Pw(\mathbf{x})$ where $Pw(.)$ denotes the power of the signal.} \citep{kereliuk2015deep} A high SNR means the reconstructed signal has low noise. Table~\ref{table:Comp2} summarizes the results on four spectrogram representations averaged over 5k generated spectrogram at 10 different runs for each dataset.
Doubling the resolution to 256$\times$256 improves both the IS and the FID, however roughly doubles the number of training parameters. Adding this note to the truncation trade-off, we can conclude that for synthesizing high quality sounds, we need to tune three major hyperparameters: $\alpha$, large batch sizes, and high resolutions. Table~\ref{table:Comp2} also shows the results of an ablation study on UrbanSound8K and the conclusions are similar to those on MCV. 

\begin{table}[t]
\centering
\footnotesize
\begin{tabular}{c|c c c c c}
\hline
Dataset               & Model                                                                        & Res. & IS                                                                                                               & FID                                                                                                               & SNR$_{db}$     \\ \hline \hline

\multirow{5}{*}{MCV}  & \begin{tabular}[c]{@{}c@{}}SA-GAN\\ (baseline)\end{tabular}                  & 128  & \begin{tabular}[c]{@{}c@{}}$-$\\ \end{tabular}   & \begin{tabular}[c]{@{}c@{}}$59.14 \pm 2.24$\\ ($\downarrow 4.54, \downarrow 1.46, -$)\end{tabular}  & $41.91$ \\ \cline{2-6} 
                      & BigGAN                                                                       & 128  & \begin{tabular}[c]{@{}c@{}}$-$\\ \end{tabular}   & \begin{tabular}[c]{@{}c@{}}$54.95 \pm 1.04$\\ ($\uparrow 2.14, \downarrow 3.48, -$)\end{tabular}      & $46.93$ \\ \cline{2-6} 
                      & \begin{tabular}[c]{@{}c@{}}BigGAN\\ (+DFN)\end{tabular}                        & 128  & \begin{tabular}[c]{@{}c@{}}$35.57 \pm 2.92$\\ ($\uparrow 1.02, \uparrow 3.46, -$)\end{tabular}       & \begin{tabular}[c]{@{}c@{}}$51.01 \pm 2.18$\\ ($\downarrow 3.52, \downarrow 2.11,  -$)\end{tabular}    & $53.72$ \\ \cline{2-6} 
                      & \multirow{2}{*}{\begin{tabular}[c]{@{}c@{}}Proposed\\ (+DFN)\end{tabular}} & 128  & \begin{tabular}[c]{@{}c@{}}$50.79 \pm 0.14$\\ ($\downarrow 3.30, \downarrow 2.48, -$)\end{tabular}   & \begin{tabular}[c]{@{}c@{}}$43.37 \pm 1.65$\\ ($\downarrow 6.88, \downarrow 4.99, -$)\end{tabular}  & $56.91$ \\ \cline{3-6} 
                      &                                                                              & 256  & \begin{tabular}[c]{@{}c@{}}$\mathbf{56.20 \pm 1.73}$\\ ($\downarrow 4.03, \downarrow 2.95, -$)\end{tabular} & \begin{tabular}[c]{@{}c@{}}$\mathbf{36.31 \pm 3.63}$\\ ($\downarrow 7.59, \downarrow 4.36, \uparrow 23.48$)\end{tabular} & $\mathbf{62.89}$ \\ \hline\hline

\multirow{4}{*}{US8K} & \begin{tabular}[c]{@{}c@{}}SA-GAN\\ (baseline)\end{tabular}                  & 256  & \begin{tabular}[c]{@{}c@{}}$34.72 \pm 0.23$\\ ($\uparrow 1.17, \uparrow 2.39, -$)\end{tabular}       & \begin{tabular}[c]{@{}c@{}} $-$\\ \end{tabular}      & $34.12$ \\ \cline{2-6} 
                      & BigGAN                                                                       & 256  & \begin{tabular}[c]{@{}c@{}}$39.16 \pm 5.07$\\ ($\downarrow 4.56, \downarrow 3.94, -$)\end{tabular}   & \begin{tabular}[c]{@{}c@{}}$37.26 \pm 3.53$\\ ($\downarrow 5.21, \downarrow 2.75, \uparrow 16.18$)\end{tabular}  & $48.14$ \\ \cline{2-6} 
                      & \begin{tabular}[c]{@{}c@{}}BigGAN\\ (+DFN)\end{tabular}                        & 256  & \begin{tabular}[c]{@{}c@{}}$48.62 \pm 0.23$\\ ($-, \uparrow 2.39, -$)\end{tabular}       & \begin{tabular}[c]{@{}c@{}}$25.63 \pm 2.42$\\ ($\uparrow 2.17, \uparrow 0.94, -$)\end{tabular}      & $55.17$ \\ \cline{2-6} 
                      & \begin{tabular}[c]{@{}c@{}}Proposed\\ (+DFN)\end{tabular}                  & 256  & \begin{tabular}[c]{@{}c@{}}$\mathbf{57.91 \pm 2.19}$\\ ($\downarrow 7.48, \downarrow 4.88, \downarrow 15.34$)\end{tabular}   & \begin{tabular}[c]{@{}c@{}}$\mathbf{16.33 \pm 1.87}$\\ ($\downarrow 3.29, \downarrow 4.88, -$)\end{tabular}  & $\mathbf{63.24}$ \\ \hline

\end{tabular}
\caption{Comparison of average IS, FID and SNR for different models trained on logarithmic DWT spectrograms. Outperforming values are in bold. Spectral normalization applied only on $D(\mathbf{x}; \theta_{d})$ for all the BigGANs and its modified variant (the proposed architecture). Res. stands for resolution.}
\label{table:Comp2}
\end{table}

\section{Conclusion}
In this paper, we have proposed a conditioning trick for the generator network based on the difference DFN measure in the spectral domain of Schur decomposition. We have experimentally demonstrated its positive impact in improving model stability at larger iterations for both baseline self-attention and on a slightly modified BigGAN architecture for class-conditional learning. We have also shown that our training scenario makes the model amenable to truncation and this helps to make a trade-off between spectrogram quality and variety. According to three objective metrics of IS, FID, and SNR, models conditioned with DFN outperform baselines in generating high quality spectrograms and less noisy reconstructed audio and speech signals.

\bibliographystyle{iclr2021_conference}

\begin{thebibliography}{79}
\providecommand{\natexlab}[1]{#1}
\providecommand{\url}[1]{\texttt{#1}}
\expandafter\ifx\csname urlstyle\endcsname\relax
  \providecommand{\doi}[1]{doi: #1}\else
  \providecommand{\doi}{doi: \begingroup \urlstyle{rm}\Url}\fi

\bibitem[Abadi et~al.(2016)Abadi, Barham, Chen, Chen, Davis, Dean, Devin,
  Ghemawat, Irving, Isard, et~al.]{abadi2016tensorflow}
Mart{\'\i}n Abadi, Paul Barham, Jianmin Chen, Zhifeng Chen, Andy Davis, Jeffrey
  Dean, Matthieu Devin, Sanjay Ghemawat, Geoffrey Irving, Michael Isard, et~al.
\newblock Tensorflow: A system for large-scale machine learning.
\newblock In \emph{12th $\{$USENIX$\}$ symposium on operating systems design
  and implementation ($\{$OSDI$\}$ 16)}, pp.\  265--283, 2016.

\bibitem[{Agbinya}(1996)]{M608394}
J.~I. {Agbinya}.
\newblock Discrete wavelet transform techniques in speech processing.
\newblock In \emph{Proceedings of Digital Processing Applications (TENCON
  '96)}, volume~2, pp.\  514--519 vol.2, 1996.

\bibitem[Alt et~al.(1995)Alt, Gr{\"a}f, Harney, Hofferbert, Lengeler, Richter,
  Schardt, and Weidenm{\"u}ller]{alt1995gaussian}
H~Alt, H-D Gr{\"a}f, HL~Harney, R~Hofferbert, H~Lengeler, A~Richter, P~Schardt,
  and HA~Weidenm{\"u}ller.
\newblock Gaussian orthogonal ensemble statistics in a microwave stadium
  billiard with chaotic dynamics: Porter-thomas distribution and algebraic
  decay of time correlations.
\newblock \emph{Physical review letters}, 74\penalty0 (1):\penalty0 62, 1995.

\bibitem[Arjovsky \& Bottou(2017)Arjovsky and Bottou]{arjovsky2017towards}
Martin Arjovsky and L{\'e}on Bottou.
\newblock Towards principled methods for training generative adversarial
  networks. arxiv e-prints, art.
\newblock \emph{arXiv preprint arXiv:1701.04862}, 2017.

\bibitem[Arjovsky et~al.(2017)Arjovsky, Chintala, and
  Bottou]{arjovsky2017wasserstein}
Martin Arjovsky, Soumith Chintala, and L{\'e}on Bottou.
\newblock Wasserstein gan.
\newblock \emph{arXiv preprint arXiv:1701.07875}, 2017.

\bibitem[Bahoura \& Rouat(2001)Bahoura and Rouat]{bahoura2001wavelet}
Mohammed Bahoura and Jean Rouat.
\newblock Wavelet speech enhancement based on the teager energy operator.
\newblock \emph{IEEE signal processing letters}, 8\penalty0 (1):\penalty0
  10--12, 2001.

\bibitem[Bollepalli et~al.(2019)Bollepalli, Juvela, and
  Alku]{bollepalli2019generative}
Bajibabu Bollepalli, Lauri Juvela, and Paavo Alku.
\newblock Generative adversarial network-based glottal waveform model for
  statistical parametric speech synthesis.
\newblock \emph{arXiv preprint arXiv:1903.05955}, 2019.

\bibitem[Brock et~al.(2017)Brock, Lim, Ritchie, and Weston]{brock2017neural}
Andrew Brock, Theodore Lim, James~M Ritchie, and Nick Weston.
\newblock Cneural photo editing with introspective adversarial networks.
\newblock In \emph{International conference on machine learning}, 2017.

\bibitem[Brock et~al.(2019)Brock, Donahue, and Simonyan]{brock2018large}
Andrew Brock, Jeff Donahue, and Karen Simonyan.
\newblock Large scale {GAN} training for high fidelity natural image synthesis.
\newblock In \emph{International Conference on Learning Representations}, 2019.
\newblock URL \url{https://openreview.net/forum?id=B1xsqj09Fm}.

\bibitem[Che et~al.(2016)Che, Li, Jacob, Bengio, and Li]{che2016mode}
Tong Che, Yanran Li, Athul~Paul Jacob, Yoshua Bengio, and Wenjie Li.
\newblock Mode regularized generative adversarial networks.
\newblock \emph{arXiv preprint arXiv:1612.02136}, 2016.

\bibitem[Chehrehsa \& Moir(2016)Chehrehsa and Moir]{chehrehsa2016speech}
Sarang Chehrehsa and Tom~James Moir.
\newblock Speech enhancement using maximum a-posteriori and gaussian mixture
  models for speech and noise periodogram estimation.
\newblock \emph{Computer Speech \& Language}, 36:\penalty0 58--71, 2016.

\bibitem[Chen et~al.(2016)Chen, Duan, Houthooft, Schulman, Sutskever, and
  Abbeel]{chen2016infogan}
Xi~Chen, Yan Duan, Rein Houthooft, John Schulman, Ilya Sutskever, and Pieter
  Abbeel.
\newblock Infogan: Interpretable representation learning by information
  maximizing generative adversarial nets.
\newblock In \emph{Advances in neural information processing systems}, pp.\
  2172--2180, 2016.

\bibitem[De~Vries et~al.(2017)De~Vries, Strub, Mary, Larochelle, Pietquin, and
  Courville]{de2017modulating}
Harm De~Vries, Florian Strub, J{\'e}r{\'e}mie Mary, Hugo Larochelle, Olivier
  Pietquin, and Aaron~C Courville.
\newblock Modulating early visual processing by language.
\newblock In \emph{Advances in Neural Information Processing Systems}, pp.\
  6594--6604, 2017.

\bibitem[Denton et~al.(2015)Denton, Chintala, Fergus, et~al.]{denton2015deep}
Emily~L Denton, Soumith Chintala, Rob Fergus, et~al.
\newblock Deep generative image models using a laplacian pyramid of adversarial
  networks.
\newblock In \emph{Advances in neural information processing systems}, pp.\
  1486--1494, 2015.

\bibitem[Donahue et~al.(2018)Donahue, Li, and
  Prabhavalkar]{donahue2018exploring}
Chris Donahue, Bo~Li, and Rohit Prabhavalkar.
\newblock Exploring speech enhancement with generative adversarial networks for
  robust speech recognition.
\newblock In \emph{IEEE Intl Conf on Acoustics, Speech and Signal Processing
  (ICASSP)}, pp.\  5024--5028, 2018.

\bibitem[Donahue et~al.(2016)Donahue, Kr{\"a}henb{\"u}hl, and
  Darrell]{donahue2016adversarial}
Jeff Donahue, Philipp Kr{\"a}henb{\"u}hl, and Trevor Darrell.
\newblock Adversarial feature learning.
\newblock \emph{arXiv preprint arXiv:1605.09782}, 2016.

\bibitem[Dumoulin et~al.(2016)Dumoulin, Belghazi, Poole, Mastropietro, Lamb,
  Arjovsky, and Courville]{dumoulin2016adversarially}
Vincent Dumoulin, Ishmael Belghazi, Ben Poole, Olivier Mastropietro, Alex Lamb,
  Martin Arjovsky, and Aaron Courville.
\newblock Adversarially learned inference.
\newblock \emph{arXiv preprint arXiv:1606.00704}, 2016.

\bibitem[Edelman(1993)]{edelman1993circular}
Alan Edelman.
\newblock The circular law and the probability that a random matrix has k real
  eigenvalues.
\newblock \emph{preprint}, 1993.

\bibitem[{Esmaeilpour} et~al.(2020){Esmaeilpour}, {Cardinal}, and
  {Koerich}]{esmaeilpour2019robust}
M.~{Esmaeilpour}, P.~{Cardinal}, and A.~L. {Koerich}.
\newblock A robust approach for securing audio classification against
  adversarial attacks.
\newblock \emph{IEEE Transactions on Information Forensics and Security},
  15:\penalty0 2147--2159, 2020.

\bibitem[Esmaeilpour et~al.(2020{\natexlab{a}})Esmaeilpour, Cardinal, and
  Koerich]{esmaeilpour2020sound}
Mohammad Esmaeilpour, Patrick Cardinal, and Alessandro~Lameiras Koerich.
\newblock From sound representation to model robustness.
\newblock \emph{arXiv preprint arXiv:2007.13703}, pp.\  1--12,
  2020{\natexlab{a}}.

\bibitem[Esmaeilpour et~al.(2020{\natexlab{b}})Esmaeilpour, Cardinal, and
  Koerich]{esmaeilpour2020unsupervised}
Mohammad Esmaeilpour, Patrick Cardinal, and Alessandro~Lameiras Koerich.
\newblock Unsupervised feature learning for environmental sound classification
  using weighted cycle-consistent generative adversarial network.
\newblock \emph{Applied Soft Computing}, 86:\penalty0 105912,
  2020{\natexlab{b}}.

\bibitem[Fang et~al.(2018)Fang, Yamagishi, Echizen, and
  Lorenzo-Trueba]{fang2018high}
Fuming Fang, Junichi Yamagishi, Isao Echizen, and Jaime Lorenzo-Trueba.
\newblock High-quality nonparallel voice conversion based on cycle-consistent
  adversarial network.
\newblock In \emph{IEEE Intl Conf on Acoustics, Speech and Signal Processing
  (ICASSP)}, pp.\  5279--5283, 2018.

\bibitem[Fedus et~al.(2017)Fedus, Rosca, Lakshminarayanan, Dai, Mohamed, and
  Goodfellow]{fedus2017many}
William Fedus, Mihaela Rosca, Balaji Lakshminarayanan, Andrew~M Dai, Shakir
  Mohamed, and Ian Goodfellow.
\newblock Many paths to equilibrium: Gans do not need to decrease a divergence
  at every step.
\newblock \emph{arXiv preprint arXiv:1710.08446}, 2017.

\bibitem[Glorot \& Bengio(2010)Glorot and Bengio]{glorot2010understanding}
Xavier Glorot and Yoshua Bengio.
\newblock Understanding the difficulty of training deep feedforward neural
  networks.
\newblock In \emph{Proceedings of the thirteenth international conference on
  artificial intelligence and statistics}, pp.\  249--256, 2010.

\bibitem[Golub \& Van~der Vorst(2000)Golub and Van~der
  Vorst]{golub2000eigenvalue}
Gene~H Golub and Henk~A Van~der Vorst.
\newblock Eigenvalue computation in the 20th century.
\newblock \emph{Journal of Computational and Applied Mathematics}, 123\penalty0
  (1-2):\penalty0 35--65, 2000.

\bibitem[Golub \& Van~Loan(2012)Golub and Van~Loan]{golub2012matrix}
Gene~H Golub and Charles~F Van~Loan.
\newblock \emph{Matrix computations}, volume~3.
\newblock JHU press, 2012.

\bibitem[Goodfellow et~al.(2014)Goodfellow, Pouget-Abadie, Mirza, Xu,
  Warde-Farley, Ozair, Courville, and Bengio]{goodfellow2014generative}
Ian Goodfellow, Jean Pouget-Abadie, Mehdi Mirza, Bing Xu, David Warde-Farley,
  Sherjil Ozair, Aaron Courville, and Yoshua Bengio.
\newblock Generative adversarial nets.
\newblock In \emph{Advances in neural information processing systems}, pp.\
  2672--2680, 2014.

\bibitem[Gulrajani et~al.(2017)Gulrajani, Ahmed, Arjovsky, Dumoulin, and
  Courville]{gulrajani2017improved}
Ishaan Gulrajani, Faruk Ahmed, Martin Arjovsky, Vincent Dumoulin, and Aaron~C
  Courville.
\newblock Improved training of wasserstein gans.
\newblock In \emph{Advances in Neural Information Processing Systems}, pp.\
  5767--5777, 2017.

\bibitem[Hannun et~al.(2014)Hannun, Case, Casper, Catanzaro, Diamos, Elsen,
  Prenger, Satheesh, Sengupta, Coates, et~al.]{hannun2014deep}
Awni Hannun, Carl Case, Jared Casper, Bryan Catanzaro, Greg Diamos, Erich
  Elsen, Ryan Prenger, Sanjeev Satheesh, Shubho Sengupta, Adam Coates, et~al.
\newblock Deep speech: Scaling up end-to-end speech recognition.
\newblock \emph{arXiv preprint arXiv:1412.5567}, 2014.

\bibitem[He et~al.(2016)He, Zhang, Ren, and Sun]{he2016deep}
Kaiming He, Xiangyu Zhang, Shaoqing Ren, and Jian Sun.
\newblock Deep residual learning for image recognition.
\newblock In \emph{Proc. IEEE conference on computer vision and pattern
  recognition}, pp.\  770--778, 2016.

\bibitem[Heusel et~al.(2017)Heusel, Ramsauer, Unterthiner, Nessler, and
  Hochreiter]{heusel2017gans}
Martin Heusel, Hubert Ramsauer, Thomas Unterthiner, Bernhard Nessler, and Sepp
  Hochreiter.
\newblock Gans trained by a two time-scale update rule converge to a local nash
  equilibrium.
\newblock In \emph{Advances in neural information processing systems}, pp.\
  6626--6637, 2017.

\bibitem[{Hossan} et~al.(2010){Hossan}, {Memon}, and {Gregory}]{M5709752}
M.~A. {Hossan}, S.~{Memon}, and M.~A. {Gregory}.
\newblock A novel approach for mfcc feature extraction.
\newblock In \emph{2010 4th International Conference on Signal Processing and
  Communication Systems}, pp.\  1--5, 2010.

\bibitem[Hu et~al.(2018)Hu, Tan, and Qian]{hu2018generative}
Hu~Hu, Tian Tan, and Yanmin Qian.
\newblock Generative adversarial networks based data augmentation for noise
  robust speech recognition.
\newblock In \emph{IEEE Intl Conf on Acoustics, Speech and Signal Processing
  (ICASSP)}, pp.\  5044--5048, 2018.

\bibitem[Karras et~al.(2018)Karras, Aila, Laine, and
  Lehtinen]{karras2018progressive}
Tero Karras, Timo Aila, Samuli Laine, and Jaakko Lehtinen.
\newblock Progressive growing of {GAN}s for improved quality, stability, and
  variation.
\newblock In \emph{International Conference on Learning Representations}, 2018.
\newblock URL \url{https://openreview.net/forum?id=Hk99zCeAb}.

\bibitem[Kereliuk et~al.(2015)Kereliuk, Sturm, and Larsen]{kereliuk2015deep}
Corey Kereliuk, Bob~L Sturm, and Jan Larsen.
\newblock Deep learning and music adversaries.
\newblock \emph{IEEE Transactions on Multimedia}, 17\penalty0 (11):\penalty0
  2059--2071, 2015.

\bibitem[Kingma \& Ba(2014)Kingma and Ba]{kingma2014adam}
Diederik~P Kingma and Jimmy Ba.
\newblock Adam: A method for stochastic optimization.
\newblock \emph{arXiv preprint arXiv:1412.6980}, 2014.

\bibitem[Kingma \& Welling(2014)Kingma and Welling]{kingma2014stochastic}
Diederik~P Kingma and Max Welling.
\newblock Stochastic gradient vb and the variational auto-encoder.
\newblock In \emph{2nd Intl Conf on Learning Representations, ICLR}, volume~19,
  2014.

\bibitem[Kodali et~al.(2017)Kodali, Abernethy, Hays, and
  Kira]{kodali2017convergence}
Naveen Kodali, Jacob Abernethy, James Hays, and Zsolt Kira.
\newblock On convergence and stability of gans.
\newblock \emph{arXiv preprint arXiv:1705.07215}, 2017.

\bibitem[Koerich et~al.(2020)Koerich, Esmaeilpour, Abdoli, Jr., and
  Koerich]{karl2020}
K.~M. Koerich, M.~Esmaeilpour, S.~Abdoli, A.~S.~Britto Jr., and A.~L. Koerich.
\newblock Cross-representation transferability of adversarial attacks: From
  spectrograms to audio waveforms.
\newblock In \emph{IEEE Intl J Conf on Neural Networks}, pp.\  1--8, 2020.

\bibitem[Krawczyk \& Gerkmann(2014)Krawczyk and Gerkmann]{krawczyk2014stft}
Martin Krawczyk and Timo Gerkmann.
\newblock Stft phase reconstruction in voiced speech for an improved
  single-channel speech enhancement.
\newblock \emph{IEEE/ACM Transactions on Audio, Speech, and Language
  Processing}, 22\penalty0 (12):\penalty0 1931--1940, 2014.

\bibitem[Kumar et~al.(2019)Kumar, Kumar, de~Boissiere, Gestin, Teoh, Sotelo,
  de~Br{\'e}bisson, Bengio, and Courville]{kumar2019melgan}
Kundan Kumar, Rithesh Kumar, Thibault de~Boissiere, Lucas Gestin, Wei~Zhen
  Teoh, Jose Sotelo, Alexandre de~Br{\'e}bisson, Yoshua Bengio, and Aaron~C
  Courville.
\newblock Melgan: Generative adversarial networks for conditional waveform
  synthesis.
\newblock In \emph{Advances in Neural Information Processing Systems}, pp.\
  14910--14921, 2019.

\bibitem[Lang \& Forinash(1998)Lang and Forinash]{lang1998time}
W~Christopher Lang and Kyle Forinash.
\newblock Time-frequency analysis with the continuous wavelet transform.
\newblock \emph{American journal of physics}, 66\penalty0 (9):\penalty0
  794--797, 1998.

\bibitem[Larsen et~al.(2015)Larsen, S{\o}nderby, Larochelle, and
  Winther]{larsen2015autoencoding}
Anders Boesen~Lindbo Larsen, S{\o}ren~Kaae S{\o}nderby, Hugo Larochelle, and
  Ole Winther.
\newblock Autoencoding beyond pixels using a learned similarity metric.
\newblock \emph{arXiv preprint arXiv:1512.09300}, 2015.

\bibitem[Leeb \& Henk(1989)Leeb and Henk]{leeb1989simultaneous}
F~Leeb and T~Henk.
\newblock Simultaneous amplitude and phase approximation for fir filters.
\newblock \emph{International journal of circuit theory and applications},
  17\penalty0 (3):\penalty0 363--374, 1989.

\bibitem[Lim \& Ye(2017)Lim and Ye]{lim2017geometric}
Jae~Hyun Lim and Jong~Chul Ye.
\newblock Geometric gan.
\newblock \emph{arXiv preprint arXiv:1705.02894}, 2017.

\bibitem[Mallat(1999)]{mallat1999wavelet}
St{\'e}phane Mallat.
\newblock \emph{A wavelet tour of signal processing}.
\newblock Elsevier, 1999.

\bibitem[Mao et~al.(2017)Mao, Li, Xie, Lau, Wang, and Smolley]{mao2017least}
Xudong Mao, Qing Li, Haoran Xie, Raymond~YK Lau, Zhen Wang, and Stephen~Paul
  Smolley.
\newblock Least squares generative adversarial networks.
\newblock In \emph{IEEE Intl Conf onComputer Vision (ICCV)}, pp.\  2813--2821,
  2017.

\bibitem[Marchesi(2017)]{marchesi2017megapixel}
Marco Marchesi.
\newblock Megapixel size image creation using generative adversarial networks.
\newblock \emph{arXiv preprint arXiv:1706.00082}, 2017.

\bibitem[Mescheder et~al.(2018)Mescheder, Geiger, and
  Nowozin]{mescheder2018training}
Lars Mescheder, Andreas Geiger, and Sebastian Nowozin.
\newblock Which training methods for gans do actually converge?
\newblock \emph{arXiv preprint arXiv:1801.04406}, 2018.

\bibitem[Mirza \& Osindero(2014)Mirza and Osindero]{mirza2014conditional}
Mehdi Mirza and Simon Osindero.
\newblock Conditional generative adversarial nets.
\newblock \emph{arXiv preprint arXiv:1411.1784}, 2014.

\bibitem[Miyato \& Koyama(2018)Miyato and Koyama]{miyato2018cgans}
Takeru Miyato and Masanori Koyama.
\newblock c{GAN}s with projection discriminator.
\newblock In \emph{International Conference on Learning Representations}, 2018.
\newblock URL \url{https://openreview.net/forum?id=ByS1VpgRZ}.

\bibitem[Miyato et~al.(2018)Miyato, Kataoka, Koyama, and
  Yoshida]{miyato2018spectral}
Takeru Miyato, Toshiki Kataoka, Masanori Koyama, and Yuichi Yoshida.
\newblock Spectral normalization for generative adversarial networks.
\newblock In \emph{International Conference on Learning Representations}, 2018.
\newblock URL \url{https://openreview.net/forum?id=B1QRgziT-}.

\bibitem[Mulgrew(2013)]{mulgrew2013stationary}
Bernard Mulgrew.
\newblock The stationary phase approximation, time-frequency decomposition and
  auditory processing.
\newblock \emph{IEEE transactions on signal processing}, 62\penalty0
  (1):\penalty0 56--68, 2013.

\bibitem[Nowozin et~al.(2016)Nowozin, Cseke, and Tomioka]{nowozin2016f}
Sebastian Nowozin, Botond Cseke, and Ryota Tomioka.
\newblock f-gan: Training generative neural samplers using variational
  divergence minimization.
\newblock In \emph{Advances in neural information processing systems}, pp.\
  271--279, 2016.

\bibitem[Odena et~al.(2017)Odena, Olah, and Shlens]{odena2017conditional}
Augustus Odena, Christopher Olah, and Jonathon Shlens.
\newblock Conditional image synthesis with auxiliary classifier gans.
\newblock In \emph{International conference on machine learning}, pp.\
  2642--2651, 2017.

\bibitem[Panti(2008)]{panti2008multidimensional}
Giovanni Panti.
\newblock Multidimensional continued fractions and a minkowski function.
\newblock \emph{Monatshefte f{\"u}r Mathematik}, 154\penalty0 (3):\penalty0
  247--264, 2008.

\bibitem[Perez et~al.(2018)Perez, Strub, de~Vries, Dumoulin, and
  Courville]{PerezSVDC18}
Ethan Perez, Florian Strub, Harm de~Vries, Vincent Dumoulin, and Aaron~C.
  Courville.
\newblock Film: Visual reasoning with a general conditioning layer.
\newblock In Sheila~A. McIlraith and Kilian~Q. Weinberger (eds.),
  \emph{Proceedings of the Thirty-Second {AAAI} Conference on Artificial
  Intelligence, (AAAI-18), the 30th innovative Applications of Artificial
  Intelligence (IAAI-18), and the 8th {AAAI} Symposium on Educational Advances
  in Artificial Intelligence (EAAI-18), New Orleans, Louisiana, USA, February
  2-7, 2018}, pp.\  3942--3951. {AAAI} Press, 2018.
\newblock URL
  \url{https://www.aaai.org/ocs/index.php/AAAI/AAAI18/paper/view/16528}.

\bibitem[Phillips(2003)]{phillips2003interpolation}
George~M Phillips.
\newblock \emph{Interpolation and approximation by polynomials}, volume~14.
\newblock Springer Science \& Business Media, 2003.

\bibitem[Piczak(2015)]{piczak2015esc}
Karol~J Piczak.
\newblock Esc: Dataset for environmental sound classification.
\newblock In \emph{Proc. 23rd ACM international conference on Multimedia}, pp.\
   1015--1018. ACM, 2015.

\bibitem[Radford et~al.(2015)Radford, Metz, and
  Chintala]{radford2015unsupervised}
Alec Radford, Luke Metz, and Soumith Chintala.
\newblock Unsupervised representation learning with deep convolutional
  generative adversarial networks.
\newblock \emph{arXiv preprint arXiv:1511.06434}, 2015.

\bibitem[Radford et~al.(2016)Radford, Metz, and Chintala]{RadfordMC15}
Alec Radford, Luke Metz, and Soumith Chintala.
\newblock Unsupervised representation learning with deep convolutional
  generative adversarial networks.
\newblock In Yoshua Bengio and Yann LeCun (eds.), \emph{4th International
  Conference on Learning Representations, {ICLR} 2016, San Juan, Puerto Rico,
  May 2-4, 2016, Conference Track Proceedings}, 2016.
\newblock URL \url{http://arxiv.org/abs/1511.06434}.

\bibitem[Raitio et~al.(2010)Raitio, Suni, Yamagishi, Pulakka, Nurminen, Vainio,
  and Alku]{raitio2010hmm}
Tuomo Raitio, Antti Suni, Junichi Yamagishi, Hannu Pulakka, Jani Nurminen,
  Martti Vainio, and Paavo Alku.
\newblock Hmm-based speech synthesis utilizing glottal inverse filtering.
\newblock \emph{IEEE transactions on audio, speech, and language processing},
  19\penalty0 (1):\penalty0 153--165, 2010.

\bibitem[Reynolds et~al.(2000)Reynolds, Quatieri, and
  Dunn]{reynolds2000speaker}
Douglas~A Reynolds, Thomas~F Quatieri, and Robert~B Dunn.
\newblock Speaker verification using adapted gaussian mixture models.
\newblock \emph{Digital signal processing}, 10\penalty0 (1-3):\penalty0 19--41,
  2000.

\bibitem[Rezende et~al.(2014)Rezende, Mohamed, and
  Wierstra]{rezende2014stochastic}
Danilo~Jimenez Rezende, Shakir Mohamed, and Daan Wierstra.
\newblock Stochastic backpropagation and approximate inference in deep
  generative models.
\newblock \emph{arXiv preprint arXiv:1401.4082}, 2014.

\bibitem[Rioul \& Vetterli(1991)Rioul and Vetterli]{rioul1991wavelets}
Olivier Rioul and Martin Vetterli.
\newblock Wavelets and signal processing.
\newblock \emph{IEEE signal processing magazine}, 8\penalty0 (4):\penalty0
  14--38, 1991.

\bibitem[Salamon et~al.(2014)Salamon, Jacoby, and
  Bello]{Salamon:UrbanSound:ACMMM:14}
J.~Salamon, C.~Jacoby, and J.~P. Bello.
\newblock A dataset and taxonomy for urban sound research.
\newblock In \emph{22nd {ACM} Intl Conf on Multimedia}, Orlando, FL, USA, Nov.
  2014.

\bibitem[Salamon \& Bello(2017)Salamon and Bello]{salamon2017deep}
Justin Salamon and Juan~Pablo Bello.
\newblock Deep convolutional neural networks and data augmentation for
  environmental sound classification.
\newblock \emph{IEEE Signal Processing Letters}, 24\penalty0 (3):\penalty0
  279--283, 2017.

\bibitem[Salimans et~al.(2016)Salimans, Goodfellow, Zaremba, Cheung, Radford,
  and Chen]{salimans2016improved}
Tim Salimans, Ian Goodfellow, Wojciech Zaremba, Vicki Cheung, Alec Radford, and
  Xi~Chen.
\newblock Improved techniques for training gans.
\newblock In \emph{Advances in neural information processing systems}, pp.\
  2234--2242, 2016.

\bibitem[Saxe et~al.(2014)Saxe, McClelland, and Ganguli]{SaxeMG13}
Andrew~M. Saxe, James~L. McClelland, and Surya Ganguli.
\newblock Exact solutions to the nonlinear dynamics of learning in deep linear
  neural networks.
\newblock In Yoshua Bengio and Yann LeCun (eds.), \emph{2nd International
  Conference on Learning Representations, {ICLR} 2014, Banff, AB, Canada, April
  14-16, 2014, Conference Track Proceedings}, 2014.
\newblock URL \url{http://arxiv.org/abs/1312.6120}.

\bibitem[S{\o}nderby et~al.(2016)S{\o}nderby, Caballero, Theis, Shi, and
  Husz{\'a}r]{sonderby2016amortised}
Casper~Kaae S{\o}nderby, Jose Caballero, Lucas Theis, Wenzhe Shi, and Ferenc
  Husz{\'a}r.
\newblock Amortised map inference for image super-resolution.
\newblock \emph{arXiv preprint arXiv:1610.04490}, 2016.

\bibitem[Sriram et~al.(2018)Sriram, Jun, Gaur, and Satheesh]{sriram2018robust}
Anuroop Sriram, Heewoo Jun, Yashesh Gaur, and Sanjeev Satheesh.
\newblock Robust speech recognition using generative adversarial networks.
\newblock In \emph{IEEE Intl Conf on Acoustics, Speech and Signal Processing
  (ICASSP)}, pp.\  5639--5643, 2018.

\bibitem[Srivastava et~al.(2017)Srivastava, Valkov, Russell, Gutmann, and
  Sutton]{srivastava2017veegan}
Akash Srivastava, Lazar Valkov, Chris Russell, Michael~U Gutmann, and Charles
  Sutton.
\newblock Veegan: Reducing mode collapse in gans using implicit variational
  learning.
\newblock In \emph{Advances in Neural Information Processing Systems}, pp.\
  3308--3318, 2017.

\bibitem[Thanh-Tung et~al.(2019)Thanh-Tung, Tran, and
  Venkatesh]{thanh2019improving}
Hoang Thanh-Tung, Truyen Tran, and Svetha Venkatesh.
\newblock Improving generalization and stability of generative adversarial
  networks.
\newblock \emph{arXiv preprint arXiv:1902.03984}, 2019.

\bibitem[Thompson \& Thompson(1996)Thompson and
  Thompson]{thompson1996minkowski}
Anthony~C Thompson and Anthony~C Thompson.
\newblock \emph{Minkowski geometry}.
\newblock Cambridge University Press, 1996.

\bibitem[Van~Loan \& Golub(1983)Van~Loan and Golub]{van1983matrix}
C.~F. Van~Loan and G.~H. Golub.
\newblock \emph{Matrix computations}.
\newblock Johns Hopkins University Press, 1983.

\bibitem[Wu et~al.(1990)Wu, Valli{\`e}res, Sprung, et~al.]{wu1990gaussian}
Hua Wu, Michel Valli{\`e}res, Donald~WL Sprung, et~al.
\newblock Gaussian-orthogonal-ensemble level statistics in a one-dimensional
  system.
\newblock \emph{Physical Review A}, 42\penalty0 (3):\penalty0 1027, 1990.

\bibitem[Yang et~al.(2017)Yang, Xie, Chen, Lou, Zhu, Huang, and
  Li]{yang2017statistical}
Shan Yang, Lei Xie, Xiao Chen, Xiaoyan Lou, Xuan Zhu, Dongyan Huang, and
  Haizhou Li.
\newblock Statistical parametric speech synthesis using generative adversarial
  networks under a multi-task learning framework.
\newblock In \emph{IEEE Automatic Speech Recognition and Understanding Workshop
  (ASRU)}, pp.\  685--691, 2017.

\bibitem[Young(2012)]{young2012wavelet}
Randy~K Young.
\newblock \emph{Wavelet theory and its applications}, volume 189.
\newblock Springer Science \& Business Media, 2012.

\bibitem[Zhang et~al.(2019)Zhang, Goodfellow, Metaxas, and
  Odena]{zhang2019self}
Han Zhang, Ian Goodfellow, Dimitris Metaxas, and Augustus Odena.
\newblock Self-attention generative adversarial networks.
\newblock In \emph{International Conference on Machine Learning}, pp.\
  7354--7363. PMLR, 2019.

\end{thebibliography}

\appendix
\section{Appendix}
\label{appendix:A}
\begin{figure}[htpb!]
  \centering
  \includegraphics[width=\textwidth]{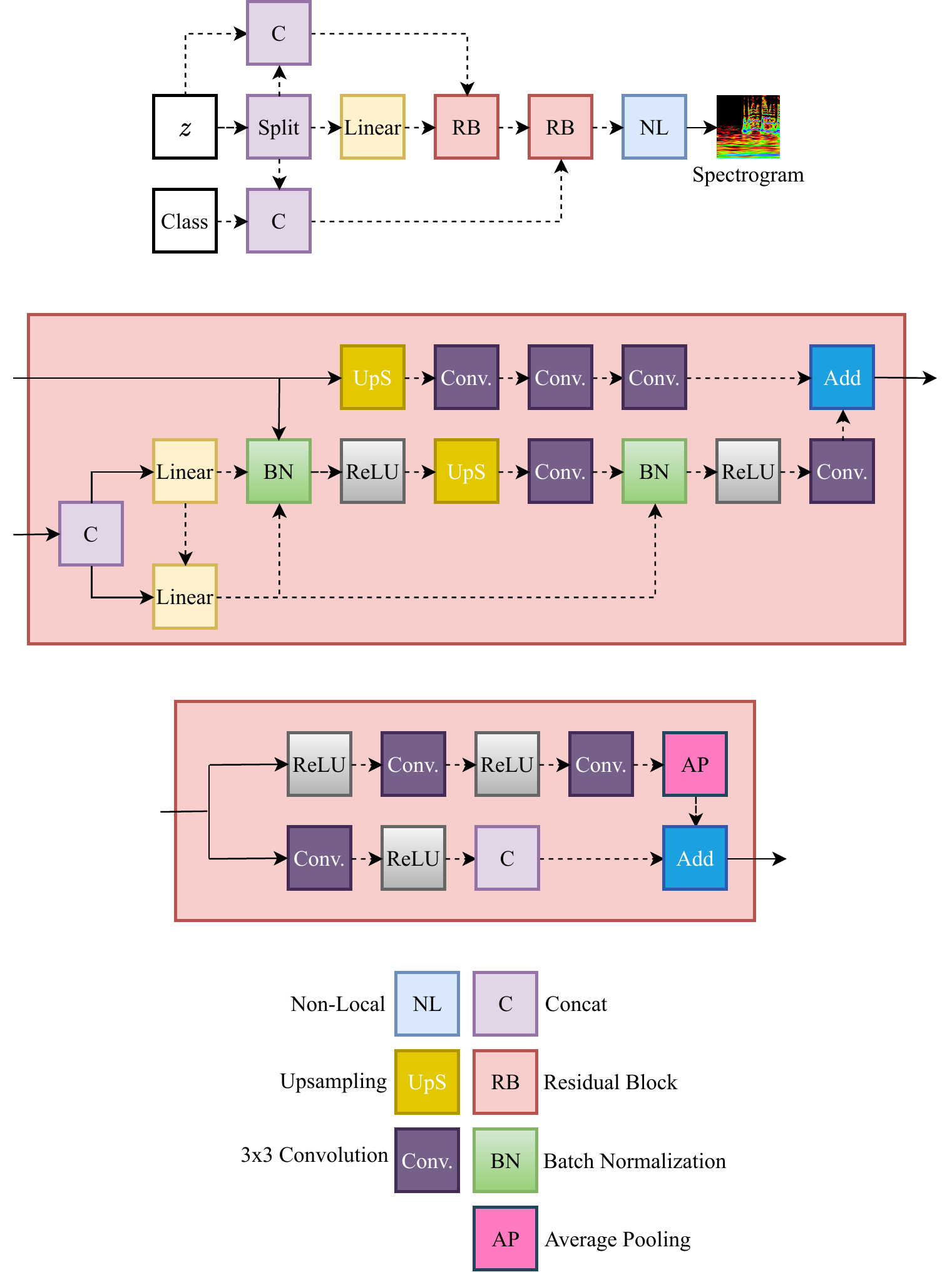}
  \caption{The proposed architecture which is the slightly modified version of BigGAN in \citep{brock2018large}. (Top): The generator architecture, (middle): a residual block in generator, (bottom): a residual block in discriminator.} 
  \label{fig:geneRes}
\end{figure}

Figure~\ref{fig:geneRes} depicts the schematic of our slightly modified version of BigGAN \citep{brock2018large}. This architecture uses the ResNet \citep{he2016deep} with different channel multipliers and shared class embeddings in the generator. Unlike the BigGAN architecture, we constantly use $3\times 3$ padded convolution with stride of 2. For the skip-$z$ connection, we use static chunks of 20-D in accordance with each residual block. Details of the networks are shown in Table~\ref{table:modifiedGAN} and Table~\ref{table:modifiedGAN2}.
\begin{table}[ht]
\centering
\small
\caption{Our slightly modified BigGAN architecture for the resolution $128 \times 128$ spectrograms. Channel stands for the width multiplier in both networks. The rest of the settings such as standing statistics for batch normalization (at the test time) are the same as \citep{brock2018large}.}
\begin{tabular}{|l|ll|l|}
\cline{1-1} \cline{4-4}
\multicolumn{1}{|c|}{\textbf{Generator}}                                & \multicolumn{1}{c}{} & \multicolumn{1}{c|}{} & \multicolumn{1}{c|}{\textbf{Discriminator}}                                       \\ \cline{1-1} \cline{4-4}
$z \in \mathbb{R}^{120} \sim \mathcal{N}(0,I)$                 &                      &                       & RGB Spectrogram $\mathbf{x}_{g} \in \mathbb{R}^{128\times 128 \times 3}$ \\ \cline{1-1} \cline{4-4} 
Linear ($20+128$)  ($\rightarrow 4\times 4 \times 16$ channel) &                      &                       & Residual Block (channel $\rightarrow 2$ channel)                         \\ \cline{1-1} \cline{4-4} 
Residual Block ($16$ channel $\rightarrow 4$ channel)         &                      &                       & Non-local Block ($64 \times 64$)                                         \\ \cline{1-1} \cline{4-4} 
Residual Block ($4$ channel $\rightarrow 1$ channel)           &                      &                       & Residual Block (channel $2\rightarrow 8$ channel)                        \\ \cline{1-1} \cline{4-4} 
Non-local Block ($128 \times 128$)                               &                      &                       & Residual Block (channel $8\rightarrow 16$ channel)                       \\ \cline{1-1} \cline{4-4} 
Batch Normalization, ReLU, Convolution                         &                      &                       & ReLU, Global Sum Pooling                                                 \\ \cline{1-1} \cline{4-4} 
$\tanh$                                                        &                      &                       & Linear $\rightarrow 1$                                                   \\ \cline{1-1} \cline{4-4} 
\end{tabular}
\label{table:modifiedGAN}
\end{table}
\begin{table}[ht]
\small
\centering
\caption{The modified BigGAN architecture for $256 \times 256$ spectrograms. This architecture has one additional residual network compared to smaller resolution $128$.}
\begin{tabular}{|l|ll|l|}
\cline{1-1} \cline{4-4}
\multicolumn{1}{|c|}{\textbf{Generator}}                                & \multicolumn{1}{c}{} & \multicolumn{1}{c|}{} & \multicolumn{1}{c|}{\textbf{Discriminator}}                                     \\ \cline{1-1} \cline{4-4} 
$z \in \mathbb{R}^{140} \sim \mathcal{N}(0,I)$                 &                      &                       & RGB Spectrogram $\mathbf{x}_{g} \in \mathbb{R}^{256\times 256 \times 3}$ \\ \cline{1-1} \cline{4-4} 
Linear ($20+128$)  ($\rightarrow 4\times 4 \times 16$ channel) &                      &                       & Residual Block (channel $\rightarrow 2$ channel)                       \\ \cline{1-1} \cline{4-4} 
Residual Block ($16$ channel $\rightarrow 4$ channel)          &                      &                       & Non-local Block ($64 \times 64$)                                       \\ \cline{1-1} \cline{4-4} 
Residual Block ($4$ channel $\rightarrow 4$ channel)           &                      &                       & Residual Block ($2$ channel $\rightarrow 4$ channel)                   \\ \cline{1-1} \cline{4-4} 
Residual Block ($4$ channel $\rightarrow 1$ channel)           &                      &                       & Residual Block ($4$ channel $\rightarrow 8$ channel)                   \\ \cline{1-1} \cline{4-4} 
Non-local Block ($128 \times 128$)                             &                      &                       & Residual Block ($8$ channel $\rightarrow 16$ channel)                  \\ \cline{1-1} \cline{4-4} 
Batch Normalization, ReLU, Convolution                         &                      &                       & ReLU, Global Sum Pooling                                               \\ \cline{1-1} \cline{4-4} 
$\tanh$                                                        &                      &                       & Linear $\rightarrow 1$                                                 \\ \cline{1-1} \cline{4-4} 
\end{tabular}
\label{table:modifiedGAN2}
\end{table}

As suggested \citep{brock2018large} for the CC conditioning in each residual block, the linear projection has been used where the bias and gain projections are centered at 0 and 1, respectively. We use Orthogonal initialization \citep{SaxeMG13} for both generator and discriminator networks. For the choice of the optimier, Adam \citep{kingma2014adam} is utilized with $\beta_{1}=0.0$ and $\beta_{2}=0.9$. The learning rate is set to $2 \cdot 10^{-4}$ and $3 \cdot 10^{-5}$ for discriminator and generator at both resolutions. The rest of settings for baselines are the same as \citep{zhang2019self} implemented in TensorFlow \citep{abadi2016tensorflow}.  

\section{Appendix}
\label{appendix:GSpecs}
We highly recommend reviewing these sources \citep{M608394,bahoura2001wavelet} about DWT representations for audio and speech signals. Then we draw attentions to the clarification of three visualizations for DWT spectrograms. Briefly, the following code snippet explains linear, logarithmic, and logarithmic real visualizations. 

\begin{lstlisting}
static float linearDWTSpectrogram(float* data)
{
    /* Generating linear visualization for DWT spectrogram
    Real part: data[0], Imaginary part: data[1]
    */
    return sqrt(data[0]*data[0]+data[1]*data[1]);
}
static float logDWTSpectrogram(float* data)
{
    /* Generating logarithmic visualization
    for DWT spectrogram */
    return log(sqrt(data[0]*data[0]+data[1]*data[1]));
}
static float logRealDWTSpectrogram(float* data)
{
    /* Generating logarithmic-real visualization
    for DWT spectrogram */
    if (data[0] == 0.0) {
        return 0.0;
    } 
    return log(abs(data[0]));
}
\end{lstlisting}

We carefully train (early stopped at checkpoints before overtraining or potential collapse) our proposed GAN architecture (the modified version of BigGAN with DFN) separately on the generated visualizations of DWT spectrograms. Afterwards, we illustrate some generated spectrograms. Additionally, we reconstruct each spectrogram into audio signals (one channel) with a predefined original phase matrix to measure relative SNR (see Figures~5-\ref{Ffig3}). 

We encourage readers listen to some reconstructed speech signals from the synthesized logarithmic DWT spectrograms using our proposed GAN architecture\footnote{$Res.=256, batch=512, ch.=96$ with spectral normalization and difference DFN on discriminator and generator, respectively.} which are included in the supplementary folder. There are six signals in this folder which their details are listed herein:
\begin{enumerate}
    \item \textbf{Original speech:} This is a random original file from MCV dataset.
    \item \textbf{Synthesized speech 1:} The reconstructed speech using the identical phase information from the aforementioned original speech. The synthesized spectrogram is generated by the model with $\epsilon=34.53$ and $\alpha=0.25$.
    \item \textbf{Synthesized speech 2:} The reconstructed speech using the identical phase information from the aforementioned original speech. The synthesized spectrogram is generated by the model with $\epsilon=47.05$ and $\alpha=0.50$.
    \item \textbf{Synthesized speech 3:} The reconstructed speech using the identical phase information from the aforementioned original speech. The synthesized spectrogram is generated by the model with $\epsilon=56.37$ and $\alpha=1$.
    \item \textbf{Synthesized speech 4:} The reconstructed speech using the identical phase information from the aforementioned original speech. The synthesized spectrogram is generated by the model with $\epsilon=78.14$ and $\alpha=2$.
    \item \textbf{Synthesized speech 5:} The reconstructed speech using the approximated phase information as proposed in \citep{krawczyk2014stft}. The synthesized spectrogram is generated by the model with $\epsilon=34.53$ and $\alpha=0.25$.
\end{enumerate}

There is (relatively) no sensible background or ambient noise in the reconstructed speech signals with synthesized spectrograms and original phase information (synthesized speech 1-4). Although, reconstructing signals with approximated phase vectors (i.e., synthesized speech 5) adds noticeable noises to the signals, still the speech is understandable. This denotes the critical role of high quality spectrogram in reconstruction.  

\begin{figure}[t!]
  \subfloat[Original phase]{\label{fig1:a}\includegraphics[width=0.42\linewidth]{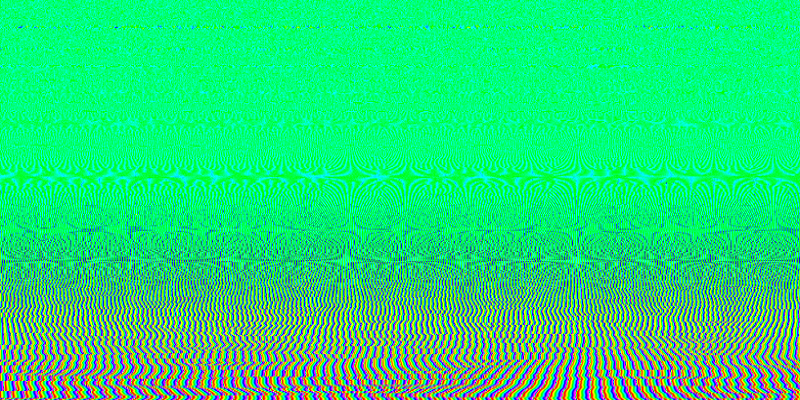}}\hfill
  \subfloat[Original Spectrogram]{\label{fig1:b}\includegraphics[width=0.42\linewidth]{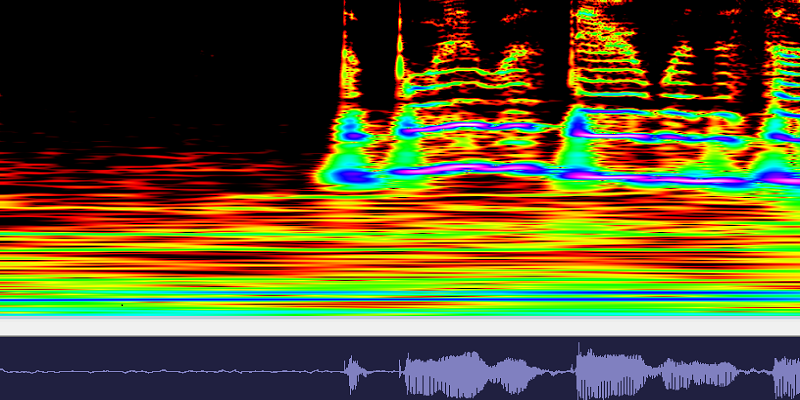}}
  \vspace{-5pt}
  \subfloat[Synthesized spectrogram ]{\label{fig1:aa}\includegraphics[width=0.42\linewidth]{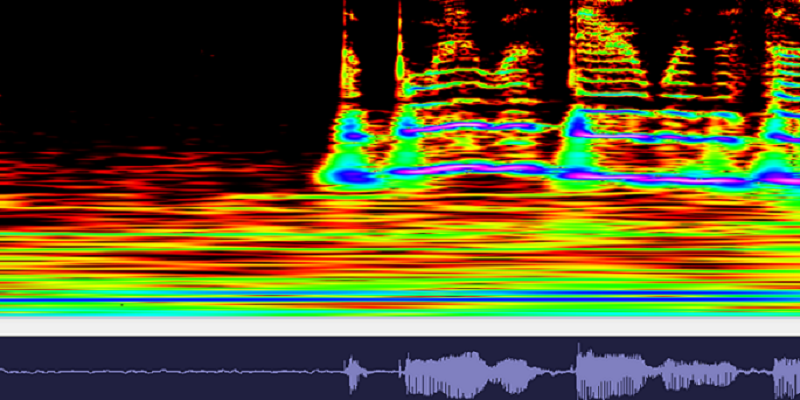}}\hfill
  \subfloat[Synthesized spectrogram]{\label{fig1:bb}\includegraphics[width=0.42\linewidth]{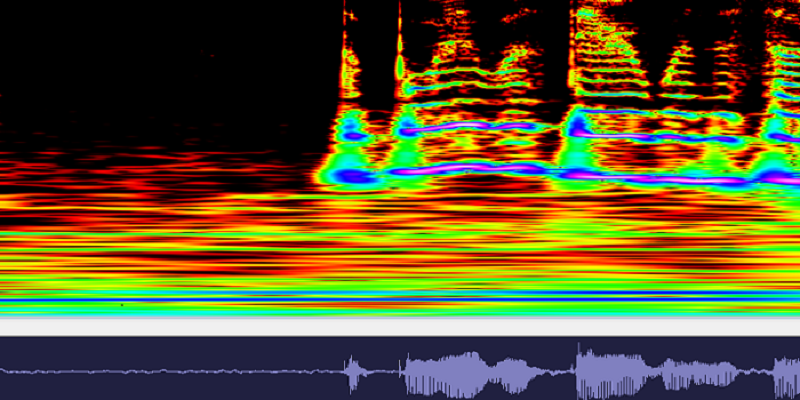}}
  \vspace{-5pt}
  \subfloat[Synthesized spectrogram]{\label{fig1:aaa}\includegraphics[width=0.42\linewidth]{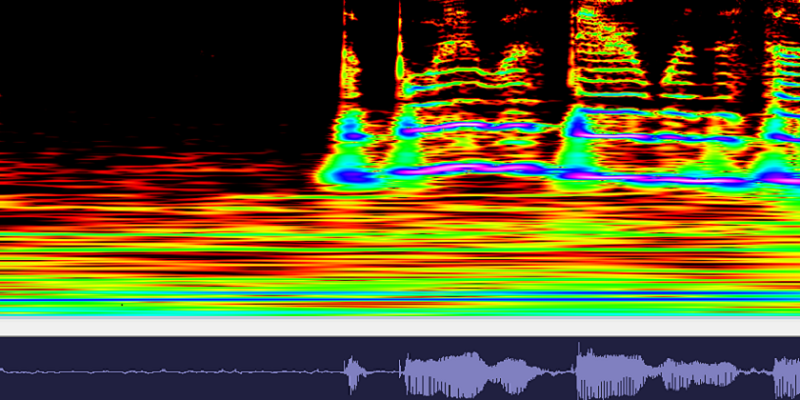}}\hfill
  \subfloat[Synthesized spectrogram]{\label{fig1:bbb}\includegraphics[width=0.42\linewidth]{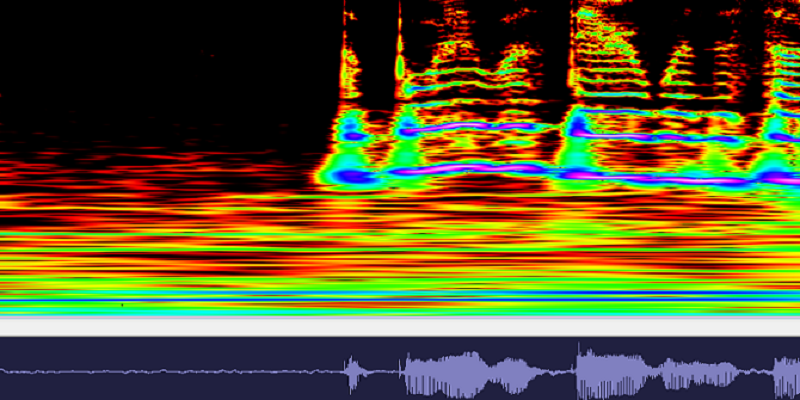}}
  \vspace{-5pt}
    \label{fFig0}
  \caption{\footnotesize Generated logarithmic DWT spectrograms with our modified BigGAN and DFN for a random signal $sig_{1}$. Synthesized signals are reconstructed with the the original phase.}
  \subfloat[Original phase]{\label{fig1:aaaa}\includegraphics[width=0.42\linewidth]{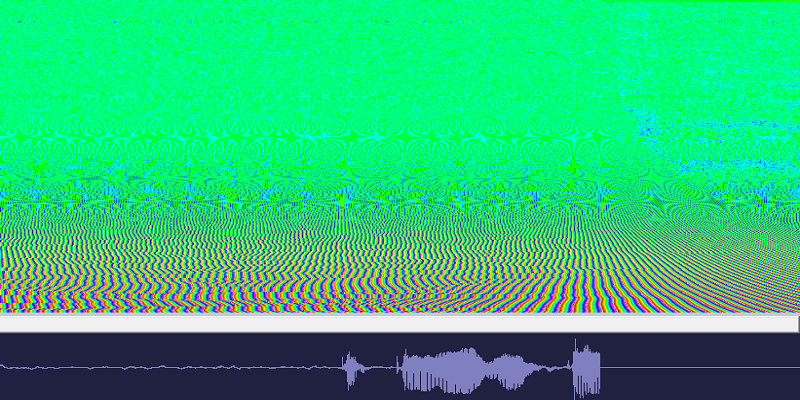}}\hfill
  \subfloat[Original spectrogram]{\label{fig1:bbbb}\includegraphics[width=0.42\linewidth]{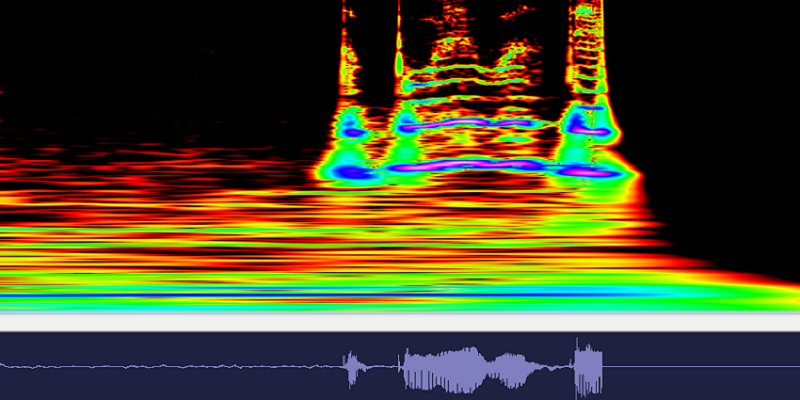}}
  \vspace{-5pt}
  \subfloat[Synthesized spectrogram]{\label{fig1:aaaaaa}\includegraphics[width=0.42\linewidth]{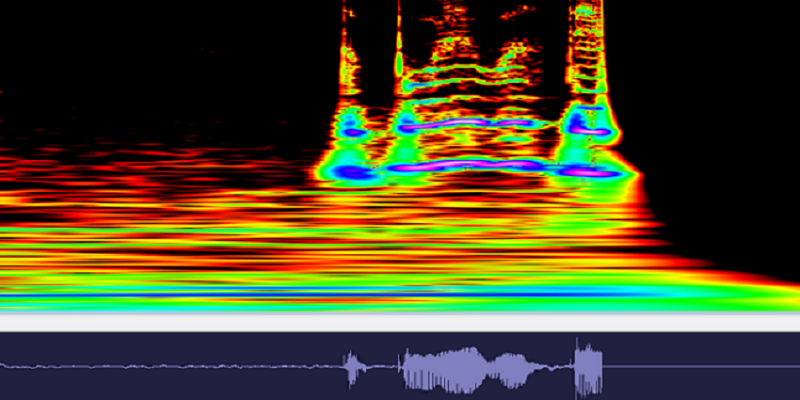}}\hfill
  \subfloat[Synthesized spectrogram]{\label{fig1:bbbbbb}\includegraphics[width=0.42\linewidth]{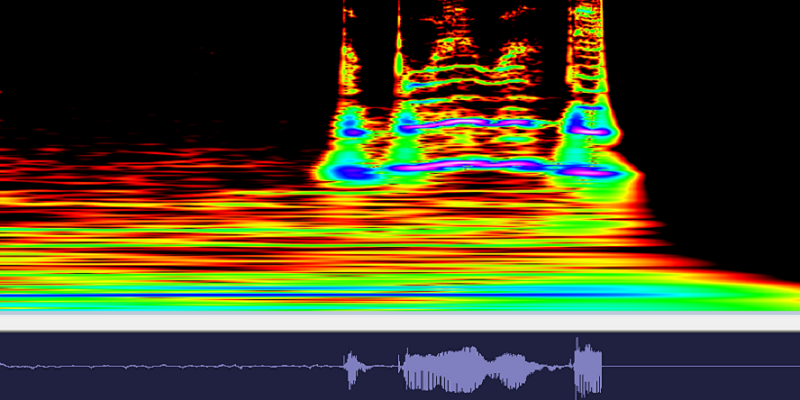}}
  \caption{\footnotesize Generated logarithmic DWT spectrograms with our modified BigGAN and DFN for another random signal $sig_{2}$. Synthesized signals are reconstructed with the the original phase.}
   \label{Ffig1}
\end{figure}

\begin{figure}[t!]
  \subfloat[Original Phase]{\label{fig1:a0}\includegraphics[width=0.42\linewidth]{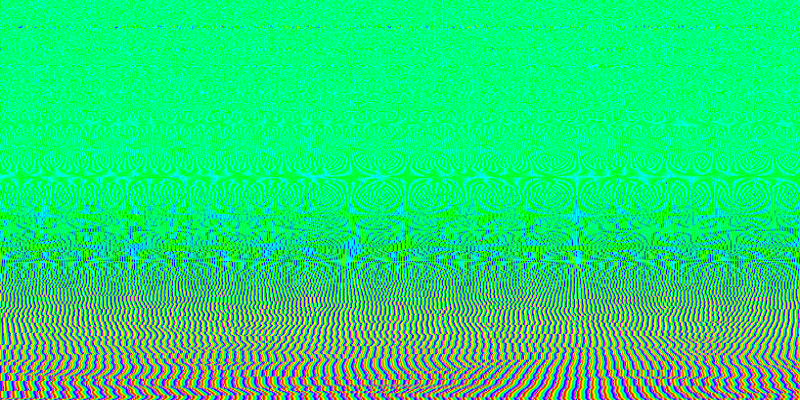}}\hfill
  \subfloat[Original spectrogram]{\label{fig1:b0}\includegraphics[width=0.42\linewidth]{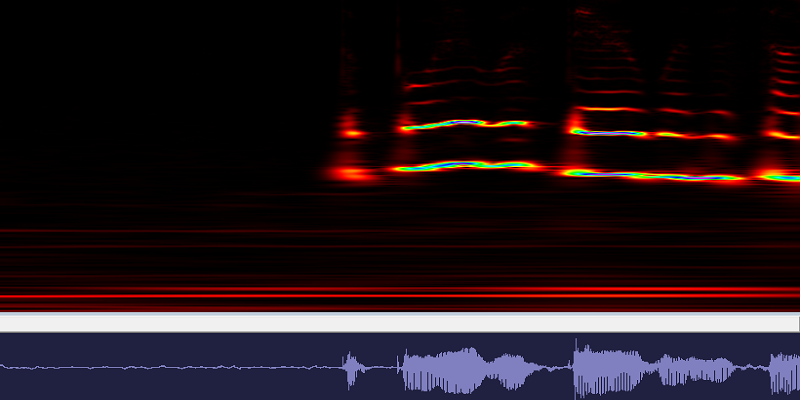}}
  \vspace{-5pt}
  \subfloat[Synthesized spectrogram]{\label{fig1:a1}\includegraphics[width=0.42\linewidth]{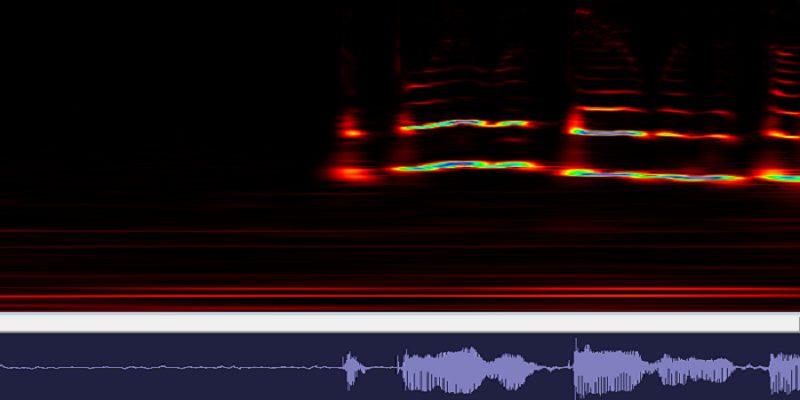}}\hfill
  \subfloat[Synthesized spectrogram]{\label{fig1:b1}\includegraphics[width=0.42\linewidth]{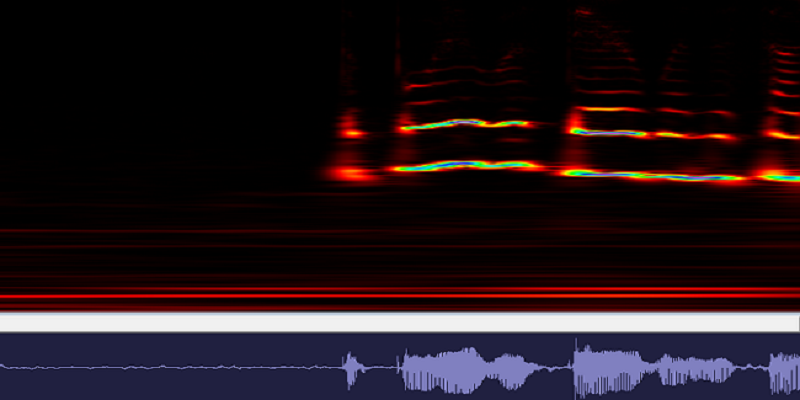}}
  \vspace{-5pt}
  \subfloat[Synthesized spectrogram]{\label{fig1:a2}\includegraphics[width=0.42\linewidth]{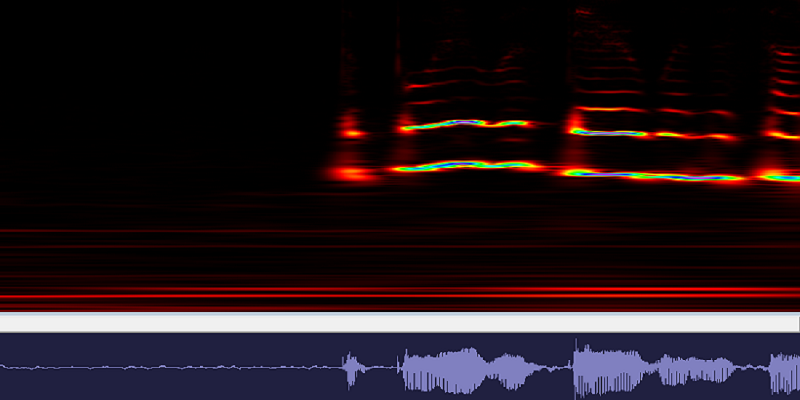}}\hfill
  \subfloat[Synthesized spectrogram]{\label{fig1:b2}\includegraphics[width=0.42\linewidth]{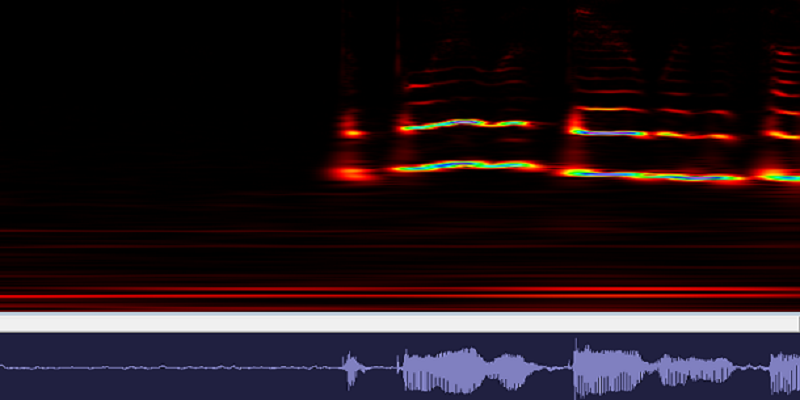}}
  \vspace{-5pt}
  \caption{\footnotesize Generated linear DWT spectrograms with our modified BigGAN and DFN for a random signal $sig_{1}$. Synthesized signals are reconstructed with the the original phase.}
  \subfloat[Original phase]{\label{fig1:a3}\includegraphics[width=0.42\linewidth]{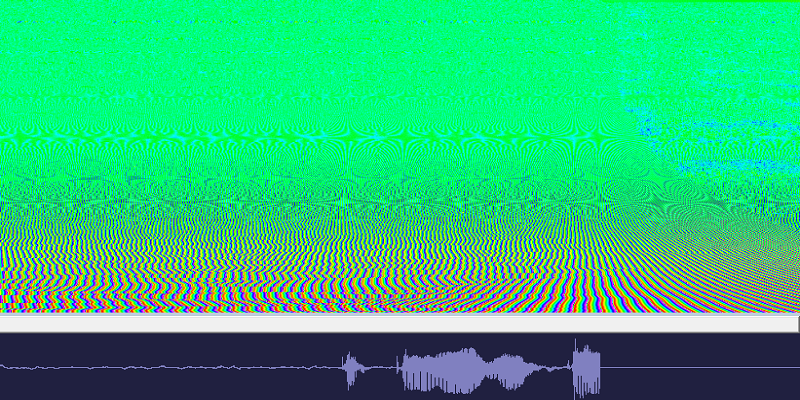}}\hfill
  \subfloat[Original spectrogram]{\label{fig1:b3}\includegraphics[width=0.42\linewidth]{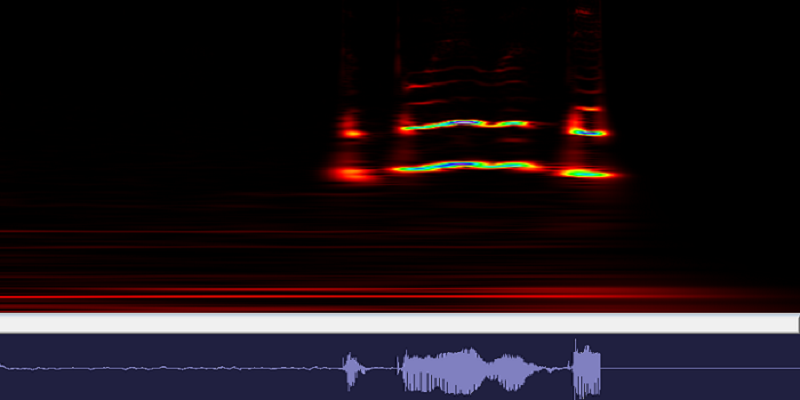}}
  \vspace{-5pt}
  \subfloat[Synthesized spectrogram]{\label{fig1:a4}\includegraphics[width=0.42\linewidth]{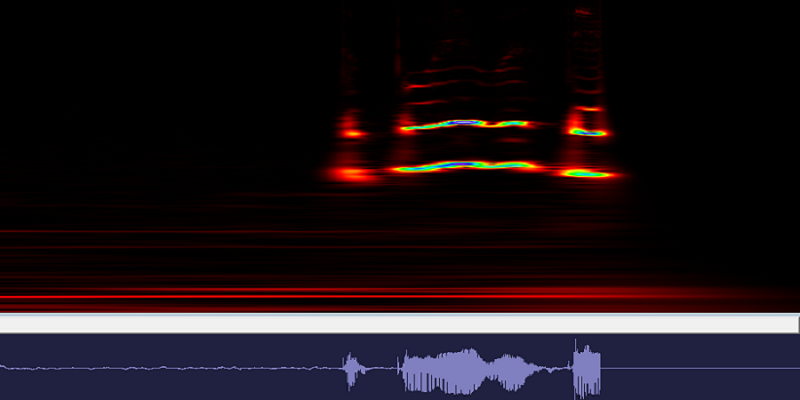}}\hfill
  \subfloat[Synthesized spectrogram]{\label{fig1:b5}\includegraphics[width=0.42\linewidth]{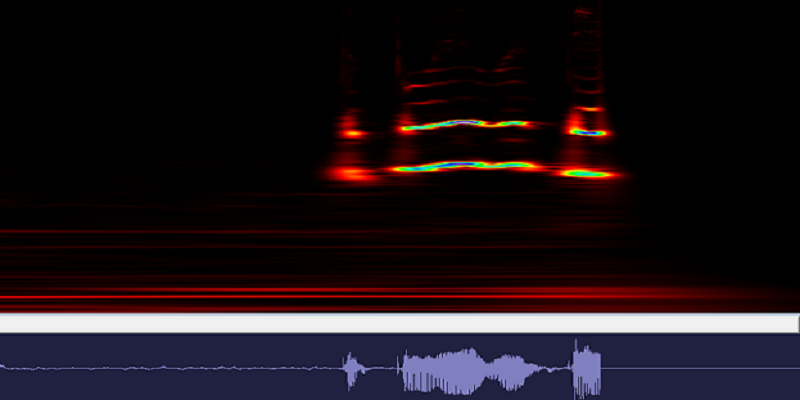}}
  \caption{\footnotesize Generated linear DWT spectrograms with our modified BigGAN and DFN for another random signal $sig_{2}$. Synthesized signals are reconstructed with the the original phase.}
  \label{Ffig2}
\end{figure}

\begin{figure}[t!]
  \subfloat[Original phase]{\label{fig1:a6}\includegraphics[width=0.42\linewidth]{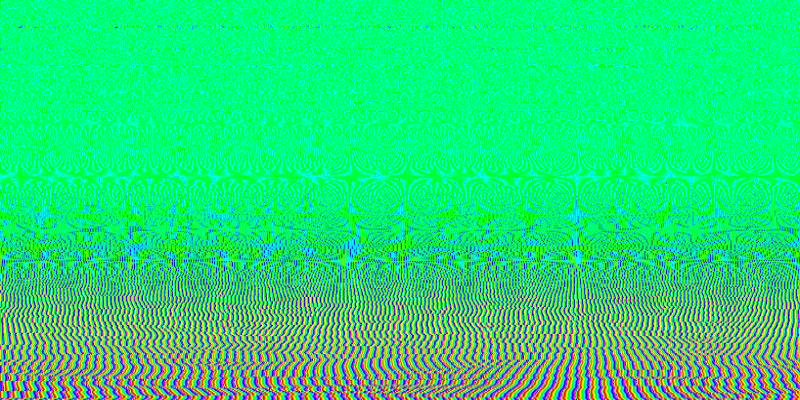}}\hfill
  \subfloat[Original spectrogram]{\label{fig1:b6}\includegraphics[width=0.42\linewidth]{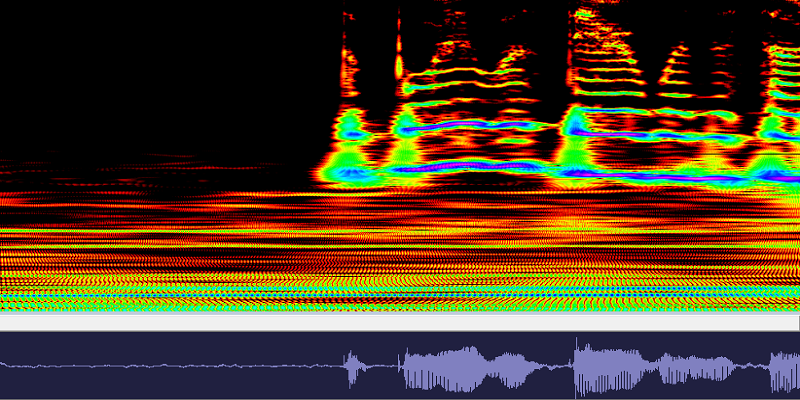}}
  \vspace{-5pt}
  \subfloat[Synthesized spectrogram]{\label{fig1:a7}\includegraphics[width=0.42\linewidth]{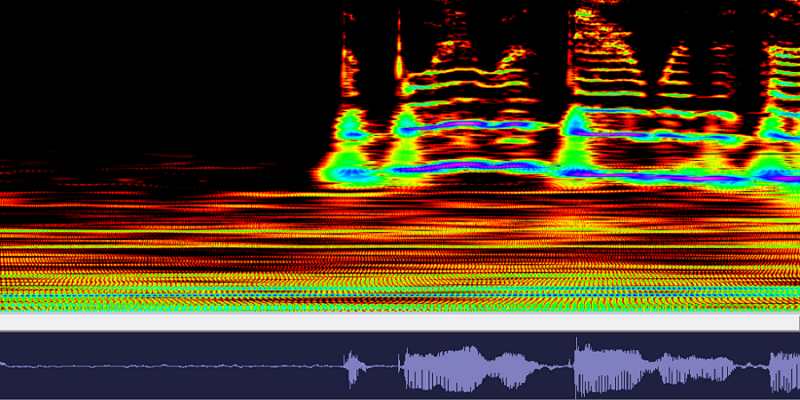}}\hfill
  \subfloat[Synthesized spectrogram]{\label{fig1:b7}\includegraphics[width=0.42\linewidth]{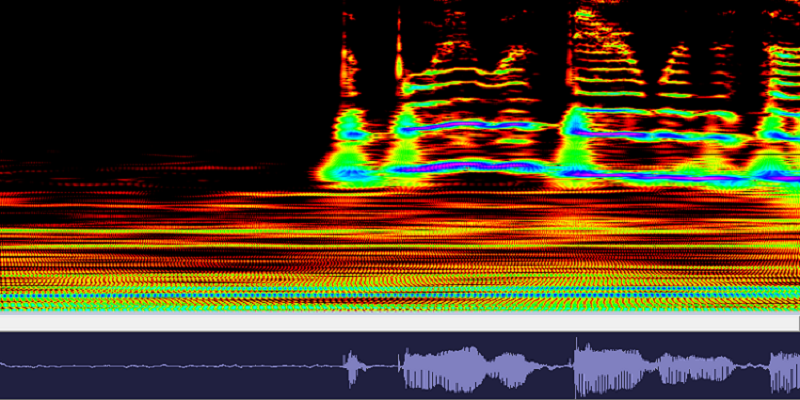}}
  \vspace{-5pt}
  \subfloat[Synthesized spectrogram]{\label{fig1:a8}\includegraphics[width=0.42\linewidth]{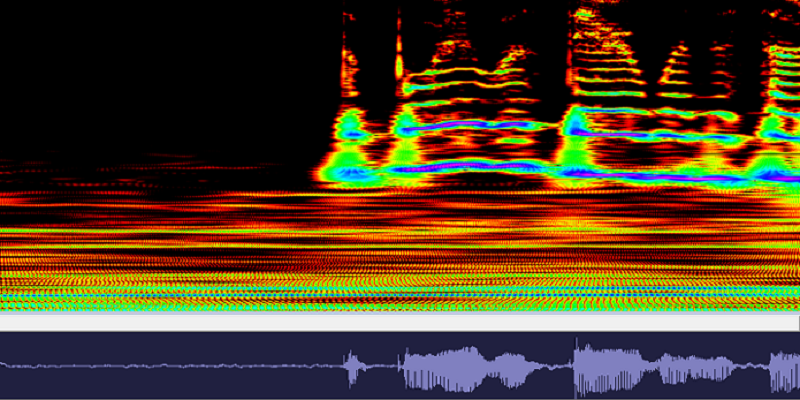}}\hfill
  \subfloat[Synthesized spectrogram]{\label{fig1:b8}\includegraphics[width=0.42\linewidth]{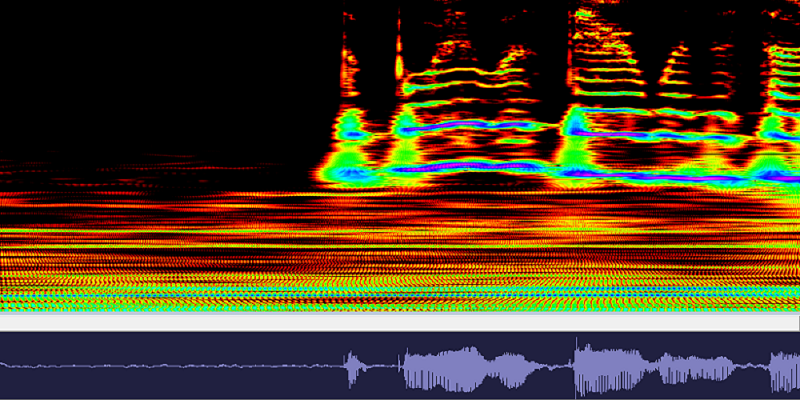}}
  \vspace{-5pt}
  \caption{\footnotesize Generated logarithmic-real DWT spectrograms with our modified BigGAN and DFN for a random signal $sig_{1}$. Synthesized signals are reconstructed with the the original phase.}
  \subfloat[Original phase]{\label{fig1:a9}\includegraphics[width=0.42\linewidth]{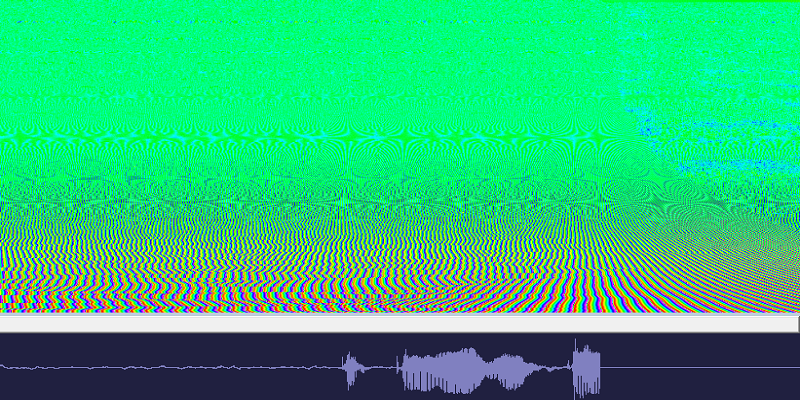}}\hfill
  \subfloat[Original spectrogram]{\label{fig1:b9}\includegraphics[width=0.42\linewidth]{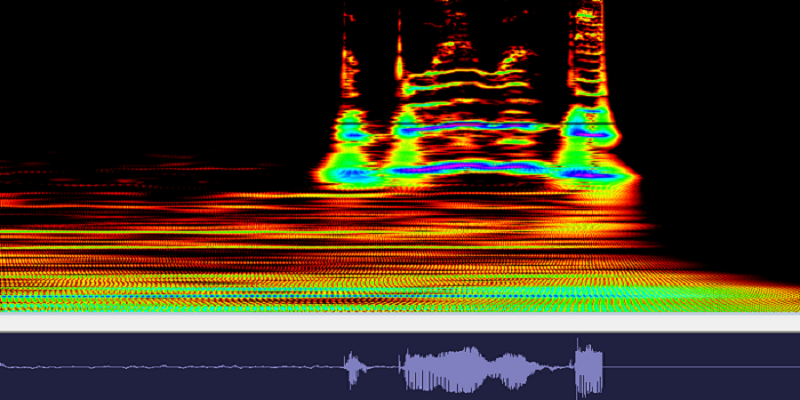}}
  \vspace{-5pt}
  \subfloat[Synthesized spectrogram]{\label{fig1:a10}\includegraphics[width=0.42\linewidth]{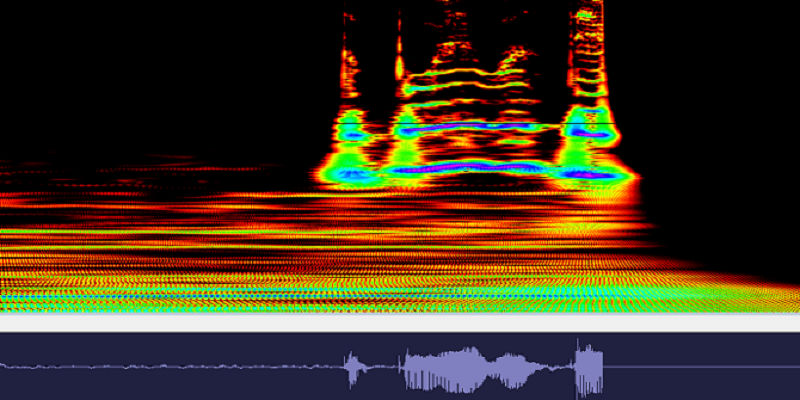}}\hfill
  \subfloat[Synthesized spectrogram]{\label{fig1:b10}\includegraphics[width=0.42\linewidth]{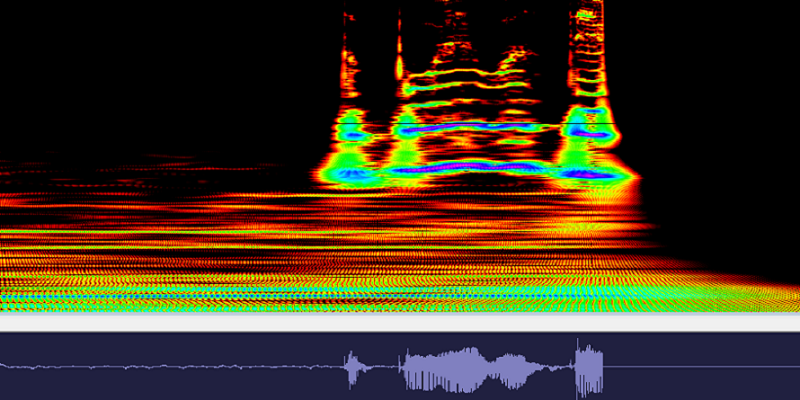}}
  \caption{\footnotesize Generated logarithmic-real DWT spectrograms with our modified BigGAN and DFN for another random signal $sig_{2}$. Synthesized signals are reconstructed with the the original phase.}
  \label{Ffig3}
\end{figure}

\end{document}